\documentclass[aps,prd,superscriptaddress,groupedaddress,nofootinbib,nobibnotes,twocolumn]{revtex4}

\usepackage{graphicx}      
                                       
\usepackage{dcolumn}
\usepackage{bm}
\usepackage{amssymb}
\usepackage{epstopdf}
\usepackage{amsmath}
\usepackage{amsfonts}
\usepackage{amsthm}
\usepackage{color}
\usepackage{mathrsfs}
\usepackage{hyperref}
\usepackage{dsfont}
\usepackage{empheq}
\usepackage{float}
\usepackage[title]{appendix}
\usepackage[]{breqn}
\usepackage{amsmath}
\usepackage{amssymb}
\usepackage{slashed}
\usepackage{todonotes}
\usepackage{comment}
\usepackage{float}
\usepackage{natbib}
\usepackage{hyperref}
\usepackage{mathtools}
\usepackage{scalerel}
\usepackage{tikz}
\usetikzlibrary{svg.path}
\definecolor{orcidlogocol}{HTML}{A6CE39}
\tikzset{
  orcidlogo/.pic={
    \fill[orcidlogocol] svg{M256,128c0,70.7-57.3,128-128,128C57.3,256,0,198.7,0,128C0,57.3,57.3,0,128,0C198.7,0,256,57.3,256,128z};
    \fill[white] svg{M86.3,186.2H70.9V79.1h15.4v48.4V186.2z}
                 svg{M108.9,79.1h41.6c39.6,0,57,28.3,57,53.6c0,27.5-21.5,53.6-56.8,53.6h-41.8V79.1z M124.3,172.4h24.5c34.9,0,42.9-26.5,42.9-39.7c0-21.5-13.7-39.7-43.7-39.7h-23.7V172.4z}
                 svg{M88.7,56.8c0,5.5-4.5,10.1-10.1,10.1c-5.6,0-10.1-4.6-10.1-10.1c0-5.6,4.5-10.1,10.1-10.1C84.2,46.7,88.7,51.3,88.7,56.8z};
  }
}
\usepackage{hyperref}  
\hypersetup{colorlinks}

\newcommand\orcidlink[1]{\href{https://orcid.org/#1}{\mbox{\scalerel*{
\begin{tikzpicture}[yscale=-1,transform shape]
\pic{orcidlogo};
\end{tikzpicture}
}{|}}}}

\newcommand{\beq}{\begin{equation}}
\newcommand{\eeq}{\end{equation}}
\newcommand{\beqa}{\begin{eqnarray}}
\newcommand{\eeqa}{\end{eqnarray}}

\usepackage{blindtext}
\begingroup\makeatletter
\catcode`,=\active
\global\let\breqn@comma,
\protected\gdef,{\ifmmode\expandafter\breqn@comma\else\expandafter\active@comma\fi}
\endgroup

\newcommand{\be}{\begin{equation}}
\newcommand{\ee}{\end{equation}}
\newcommand{\ba}{\begin{eqnarray}}
\newcommand{\ea}{\end{eqnarray}}

\newcommand{\barr}{\begin{array}}
\newcommand{\earr}{\end{array}}

\newcommand\lsim{\mathrel{\rlap{\lower4pt\hbox{\hskip1pt$\sim$}}
        \raise1pt\hbox{$<$}}}
\newcommand\gsim{\mathrel{\rlap{\lower4pt\hbox{\hskip1pt$\sim$}}
        \raise1pt\hbox{$>$}}}

\usepackage{scalerel}
\usepackage{tikz}
\usetikzlibrary{svg.path}
\definecolor{orcidlogocol}{HTML}{A6CE39}
\tikzset{
  orcidlogo/.pic={
    \fill[orcidlogocol] svg{M256,128c0,70.7-57.3,128-128,128C57.3,256,0,198.7,0,128C0,57.3,57.3,0,128,0C198.7,0,256,57.3,256,128z};
    \fill[white] svg{M86.3,186.2H70.9V79.1h15.4v48.4V186.2z}
                 svg{M108.9,79.1h41.6c39.6,0,57,28.3,57,53.6c0,27.5-21.5,53.6-56.8,53.6h-41.8V79.1z M124.3,172.4h24.5c34.9,0,42.9-26.5,42.9-39.7c0-21.5-13.7-39.7-43.7-39.7h-23.7V172.4z}
                 svg{M88.7,56.8c0,5.5-4.5,10.1-10.1,10.1c-5.6,0-10.1-4.6-10.1-10.1c0-5.6,4.5-10.1,10.1-10.1C84.2,46.7,88.7,51.3,88.7,56.8z};
  }
}

\begin{document}

\title{Minimizing Contaminant Leakage in Internal Linear Combination Maps Using a Data-Driven Approach}
\date{\today}

\author{Kristen M.~Surrao\,\orcidlink{0000-0002-7611-6179}}
\affiliation{Department of Physics, Columbia University, New York, NY 10027, USA}

\author{Shivam Pandey}
\affiliation{William H.~Miller III Department of Physics \& Astronomy, Johns Hopkins University, Baltimore, MD 21218, USA}
\affiliation{Department of Physics, Columbia University, New York, NY 10027, USA}

\author{J.~Colin Hill\,\orcidlink{0000-0002-9539-0835}}
\affiliation{Department of Physics, Columbia University, New York, NY 10027, USA}

\author{Eric J.~Baxter}
\affiliation{Institute for Astronomy, University of Hawai'i, 2680 Woodlawn Drive, Honolulu, HI 96822,
USA}

\begin{abstract}
The thermal Sunyaev-Zel'dovich (tSZ) effect, the inverse-Compton scattering of cosmic microwave background (CMB) photons off high-energy electrons, is a powerful probe of hot, ionized gas in the Universe. It is often measured via cross-correlations of CMB data with large-scale structure (LSS) tracers to constrain gas physics and improve cosmological constraints. The largest source of bias to these measurements is the leakage of poorly understood thermal dust emission from star-forming galaxies --- the cosmic infrared background (CIB) --- into the tSZ maps. This CIB contamination is difficult to clean via multifrequency component separation methods, such as internal linear combination (ILC), due to uncertainty in its spectral energy distribution (SED), which exhibits spatial and line-of-sight variation and decorrelation. Thus, improved ILC-based techniques have been developed to null (``deproject'') both the CIB and its first moment with respect to the emissivity index $\beta$ in order to robustly remove the CIB despite the lack of first-principles knowledge of its SED. While decreasing the bias, such procedures can significantly increase the noise in the resulting ILC maps. In this paper, we develop a data-driven algorithm for determining the optimal CIB SED to deproject when measuring a tSZ-LSS cross-correlation, obviating the need to deproject the first moment in the ILC map used for such a measurement. Our method gives an unbiased cross-correlation with increased signal-to-noise. We demonstrate its efficacy on simulations, finding a 60\% improvement in the signal-to-noise ratio for an example tSZ cross-correlation with a halo sample at redshifts $0.8 < z < 1.8$, as compared to moment deprojection approaches. Though used here for CIB removal in tSZ cross-correlations, our method is broadly applicable to minimizing contaminant leakage in ILC maps. Our code is publicly available in \verb|CIB-deproj|\footnote{\url{https://github.com/kmsurrao/CIB-deproj}}.
\end{abstract}

\maketitle

\section{Introduction}
\label{sec.introduction}

The thermal Sunyaev-Zel'dovich (tSZ) effect is the inverse-Compton scattering of cosmic microwave background (CMB) photons off high-energy electrons \cite{SZ1970}, which allows us to probe hot, ionized gas in the Universe. In CMB thermodynamic temperature units, the tSZ spectral energy distribution (SED) is given by
\begin{equation}
    \label{eq.tsz_sed}
    f_\nu = T_{\rm CMB} \left( x \coth{\frac{x}{2}} - 4 \right) \; \text{ with } \; x=\frac{h\nu}{k_B T_{\mathrm{CMB}}} \, ,
\end{equation}
where $h$ is Planck's constant, $k_B$ is the Boltzmann constant, $T_{\mathrm{CMB}}=2.726$ K is the CMB temperature at redshift $z=0$. Compton-$y$ maps, the frequency-independent templates of the tSZ effect, are often constructed from multi-frequency CMB data (e.g., \cite{Planck2015ymap,Madhavacheril:2019nfz, Bleem:2021ymap, Coulton2023ACT}) and then cross-correlated with galaxy surveys, weak lensing maps, or other large-scale structure (LSS) tracers to probe gas physics and improve cosmological constraints. These measurements allow us to constrain gas thermodynamics, pressure profiles, feedback, and galaxy formation physics \cite{Birkinshaw_1999, Carlstrom:2002na}.

Many past works have cross-correlated or performed stacking analyses with $y$-maps and galaxies, with the goal of constraining the halo $Y$-$M$ relation (the relationship between the integrated Compton-$y$ parameter and the mass of the halo), constraining the bias-weighted thermal gas pressure of the Universe, and/or constraining astrophysical feedback processes via the halo pressure profiles. For example, Refs.~\cite{Pandey_des:2019, Vikram2017, Hill2018sz, 2013A&A...557A..52P, Tanimura2020, Koukoufilippas:2019ilu} used \emph{Planck} $y$-maps (e.g., \cite{Planck2015ymap,Tanimura2022,McCarthy:2023hpa}) to measure cross-correlations with various galaxy samples (e.g.,~\cite{des2015, Bilicki:2014, Bilicki:2016}). Similarly, Ref.~\cite{DESSPT:2022xmu} used $y$-maps built from combined \emph{Planck} and South Pole Telescope (SPT) data, and Refs.~\cite{Vavagiakis:2021ilq, Madhavacheril:2019nfz, Amodeo:2020mmu, Schaan:2020wtv} used $y$-maps constructed from Atacama Cosmology Telescope (ACT) and \emph{Planck} data, to measure tSZ-LSS cross-correlations. Past analyses have also measured cross-correlations of $y$-maps with weak gravitational lensing maps to constrain feedback models via their signatures in the intracluster medium (ICM) pressure profile (e.g., \cite{Battaglia:2014era, VanWaerbeke:2013cfa, Hojjati:2016nbx, Osato:2019ynf, Ma:2020cir, DES:2021sgf, Gatti2021, Troster:2021gsz}). Others have measured tSZ cross-correlations with CMB lensing maps to place constraints on ICM astrophysics at higher redshifts~\cite{Hill:2013dxa, McCarthy:2023cwg}. These analyses establish the importance of tSZ cross-correlations in understanding gas thermodynamics over a large dynamical range of halo masses and redshifts and using it to constrain astrophysical feedback. It is crucial to increase the significance of these measurements to make such inferences more precise. 

In general, CMB data analyses only have access to a finite number of individual frequency maps that can be combined (e.g., via an internal linear combination \cite{Bennett2003, Tegmark2003, Eriksen2004, Delabrouille2009}) to reconstruct a tSZ map using the known SED of the tSZ effect. However, other contaminants can also leak into the maps if not explicitly nulled (``deprojected'') in the reconstruction procedure. The single largest source of astrophysical bias to such tSZ-LSS cross-correlation measurements is the leakage of poorly understood thermal dust emission from star-forming galaxies, which is the dominant signal on small scales in the mm-wave sky at high frequencies. This dust emission forms the cosmic infrared background (CIB). The CIB can result in biases on the order of tens of percent to tSZ cross-correlations without special treatment to remove it (e.g., \cite{Hill:2013dxa, Yan:2018wlf, Gatti2021, McCarthy:2023cwg}).

The CIB SED is often modeled roughly as a modified blackbody (MBB):
\begin{equation}
    \label{eq.MBB}
    g_\nu \propto \left( \frac{\nu}{\nu_0} \right)^{\beta+3} \frac{1}{e^{h\nu/(k_B T_d)}-1}  \, ,
\end{equation}
where $g_\nu$ is the SED as a function of frequency $\nu$, $\nu_0$ is the CIB pivot frequency, $\beta$ is the emissivity index, $h$ is Planck's constant, $k_B$ is the Boltzmann constant, and $T_d$ is the effective dust temperature. The pivot frequency can be arbitrarily set, but one must make some informed choice for $\beta$ and $T_d$.  Note that the overall amplitude of this SED is irrelevant for our purposes, as it cancels when deprojecting the CIB in a constrained ILC (see Eq.~\eqref{eq.cilc} below).

In reality, one does not know \emph{a priori} what values of $\beta$ and $T_d$ to use. Moreover, the CIB decorrelates across frequencies and is not accurately described by a single, fixed SED, since it is sourced by halos across a wide range of redshifts (though this is a relatively small effect across the range of frequencies used in most CMB experiments \cite{Websky_2020, Omori:2022, Lenz:2019ugy, Mak:2016ykk}). One can Taylor-expand the CIB SED in $\beta$ or $T_d$ for small fluctuations away from the fiducial MBB, where the first term is the regular MBB and the second term is the first moment, often referred to as $d\beta$ when Taylor-expanding in $\beta$:
\begin{equation}
    g_\nu(\beta) = g_\nu(\beta_0) + (\beta-\beta_0) \left.\frac{\partial g_\nu}{\partial \beta} \right|_{\beta_0} \, .
\end{equation}
The first moment SED can then be deprojected in internal linear combination (ILC) procedures, described in Sec.~\ref{sec.ILC}, to more robustly remove CIB contamination when $\beta$ is not well known or when $\beta$ varies within the analysis domain \cite{2017MNRAS.472.1195C}. However, this comes at the cost of significantly increasing the noise in the ILC $y$-map. Ref.~\cite{McCarthy:2023hpa} found that deprojecting $d\beta$ led to the auto-power spectrum of the $y$-map being five times as large as that obtained via a standard ILC procedure, due to enhanced noise leakage, translating to a loss in signal-to-noise in cross-correlation measurements involving such a $y$-map.  This is a classic example of a bias-variance trade-off: reducing bias (by deprojecting $d\beta$) leads to increased variance (noise).

If it were possible to determine an effective value of $\beta$ that optimally cleaned the CIB in an ILC $y$-map for a given cross-correlation measurement, one could avoid having to deproject the $d\beta$ SED, thus improving the signal-to-noise of the measurement while maintaining robust CIB removal. The optimal $\beta$ value will, in general, depend on the tracer with which we are cross-correlating the tSZ map. In this paper, we present such an algorithm for determining an optimal value of $\beta$ for CIB deprojection (for a given cross-correlation measurement) from the data alone, i.e., not relying on any theoretical model aside from the general form of the MBB SED. To do so, we develop a technique to effectively inflate the CIB contributions to the frequency maps, and compare cross-correlations of $y$-maps built from the original frequency maps to those built from the modified frequency maps. This procedure is fully data-driven and does not rely on the use of simulations or theory to estimate any quantities. As described in more detail in Sec.~\ref{sec.algorithm}, our method is constructed to remove portions of the CIB that are correlated with the specific tracer with which we are interested in cross-correlating the $y$-map, thus resulting in an unbiased cross-correlation. It is not designed to produce an unbiased $y$-map for general use, but rather, only for the specific cross-correlation of interest.

While this paper focuses on CIB removal for tSZ cross-correlations as a specific use case of the method described, the technique is not limited to this application. It is broadly applicable to minimizing contaminant leakage in ILC reconstruction algorithms. Examples of potential additional use cases are described later in Sec.~\ref{sec.discussion}.

The remainder of this paper is organized as follows. In Sec.~\ref{sec.ILC} we review ILC methods. In Sec.~\ref{sec.algorithm} we present the algorithm for determining the optimal effective $\beta$ value. In Sec.~\ref{sec.simulations} we describe our simulations and analysis set-up. In Sec.~\ref{sec.results} we demonstrate the efficacy of the algorithm on the simulations, and we conclude in Sec.~\ref{sec.discussion}.

\section{Internal Linear Combination}

\label{sec.ILC}

\subsection{Standard ILC}

ILC~\cite{Bennett2003, Tegmark2003, Eriksen2004, Delabrouille2009} is an approach that constructs component-separated maps by finding the minimum-variance linear combination of the observed frequency maps that satisfies the constraint of unit response to the signal of interest. In pixel space at some pixel $p$, the signal of interest can be expressed as $\hat{y}(p) = \sum_i w_i T_i(p)$, where $T_i(p)$ is the temperature fluctuation in the $i$th frequency map and $w_i$ is the weight applied to that frequency map. Then the weights are obtained by minimizing the variance of the map while preserving the response to $y$:
\be
\begin{aligned}
\min_{w^i} \quad & \sigma^2_{\hat{y}}=N^{-1}_{\mathrm{pix}} \sum_p (\hat{y}(p)-\langle y \rangle)^2\\
\textrm{s.t.} \quad & w^i f_i = 1 \, ,
\end{aligned}
\ee
where Einstein summation convention is assumed, $N_{\mathrm{pix}}$ is the number of pixels, $\langle y \rangle$ is the average signal across pixels, $f_i$ is the spectral response of the signal of interest at frequency channel $i$, and $i \in \{1,...,N_{\rm freqs}\}$ where $N_{\rm freqs}$ is the total number of frequency channels. Lagrange multipliers can be used to solve for the weights \cite{Eriksen2004}:
\be
w^i = \frac{f_j(\hat{R}^{-1})_{ij}}{f_k(\hat{R}^{-1})_{km}f_m}
\ee with
\be
\label{eq.real_space_cov}
\hat{R}_{ij}=N^{-1}_{\mathrm{pix}}\sum_p T_i(p) T_j(p) \, ,
\ee
where $\hat{R}_{ij}$ is the empirical frequency-frequency covariance matrix of the observed maps. The size of the real-space domains on which the ILC is performed must be large enough to mitigate the ``ILC bias'' that results from computing the ILC weights using a small number of modes \cite{Delabrouille2009}.

ILC can alternatively be formulated in harmonic space in what is known as ``harmonic ILC", giving $\ell$-dependent weights:
\be
\label{eq.HILCweights}
w^i_{\ell} = \frac{ \left( \hat{R}_{\ell}^{-1} \right)_{ij} f_j }{f_k \left(\hat{R}_{\ell}^{-1} \right)_{km} f_m } 
\ee 
with
\be
 \left(\hat{R}_{\ell} \right)_{ij}= \sum_{\ell' = \ell-\Delta \ell /2}^{\ell+\Delta \ell /2} \frac{2\ell'+1}{4\pi} C_{\ell'}^{ij} \, .
\ee
In harmonic ILC, the ILC bias is mitigated by making the multipole bin width $\Delta \ell$ sufficiently wide. 

We use harmonic ILC throughout the remainder of this paper, but later discuss how our method can be extended to other types of ILC procedures.

\subsection{Constrained ILC}

Standard ILC methods ensure that the signal of interest is preserved. While the leakage of other components present in the temperature maps is reduced via the variance minimization, contaminants generally still propagate into the final ILC map at some level. To mitigate such effects, one can explicitly deproject some component from the ILC map, i.e., require that the ILC weights have zero response to a contaminant with some specified SED. In harmonic ILC, and letting  $g_i$ represent the deprojected component's spectral response at the $i$th frequency channel, this constraint can be expressed as 
\begin{equation}
    \label{eq.ilc_deproj_constraint}
     w^i_{\ell} g_i = 0 \, .
\end{equation}
Variance minimization while preserving unit response to the signal of interest along with this additional constraint then yields the solutions to the weights as follows \cite{Remazeilles_2010}:
\begin{widetext}
\begin{equation}
\label{eq.cilc}
    w^j_{\ell} = \frac{ \left(g_k (\hat{R}_{\ell}^{-1})_{km} g_m \right) f_i (\hat{R}_{\ell}^{-1})_{ij} - \left(f_k (\hat{R}_{\ell}^{-1})_{km} g_m \right) g_i (\hat{R}_{\ell}^{-1})_{ij}}{\left(f_k (\hat{R}_{\ell}^{-1})_{ki} f_i\right) \left(g_m (\hat{R}_{\ell}^{-1})_{mn} g_n \right) - \left(f_k (\hat{R}_{\ell}^{-1})_{ki} g_i \right)^2} \, .
\end{equation}
\end{widetext} 
Eq.~\eqref{eq.cilc} provides the equation for the ILC weights when one component is deprojected, but one can deproject as many as $N_{\mathrm{freqs}}-1$ components. In the general case of this multiply constrained ILC, the weights can be found as in Refs.~\cite{Kusiak:2023hrz, Remazeilles:2021}. In addition to deprojecting any given component, one can deproject moments of the SED, as described in Sec.~\ref{sec.introduction}.

While adding deprojection constraints can reduce biases in the final map, it comes with the standard bias-variance tradeoff that the resulting noise in the ILC map necessarily increases, since the feasible region of the optimization is smaller. Typical moment deprojection approaches (deprojecting both the SED of a contaminant and its first moment with respect to some parameter that is not precisely known, as described in Sec.~\ref{sec.introduction}) thus result in large amounts of noise in the ILC maps. 

The optimal CIB SED that is relevant for a particular cross-correlation can be different for different tracers, since the CIB SED evolves with different redshifts and the environments of the dusty galaxies. Thus, this moment deprojection is typically what has been done in recent analyses for CIB removal in order to mitigate biases that would otherwise be significant (e.g. \cite{McCarthy:2023cwg, Efstathiou:2025ckq, pandey2025_inprep}). However, cross-correlations with ILC maps that have removed the CIB in this way often have low signal-to-noise relative to maps in which the first moment is not deprojected.

\section{Algorithm for Determining the Emissivity Index}

\label{sec.algorithm}

Our goal in this section is to find the optimal CIB SED (using a data-driven approach) to deproject during tSZ reconstruction with ILC, such that the resulting $y$-map produces an unbiased cross-correlation with a particular LSS tracer. Specifically, we fix $T_d$ to a given value and develop a procedure to find the optimal emissivity index $\beta$ (optimal in the sense of minimizing the CIB contamination to a tSZ cross-correlation with the given tracer). Since $T_d$ and $\beta$ are highly correlated, for any given $T_d$ value, a different $\beta$ is optimal. Here we fix $T_d$ to turn the problem into a simple 1D search over $\beta$. In theory, it is also possible to jointly optimize $T_d$ and $\beta$, but this is not necessary due to their degeneracy, at least within the range of frequencies used in most CMB experiments, which are in the Rayleigh-Jeans regime of the CIB SED. 

The main idea behind our approach is to alter the CIB component in the individual frequency maps, while keeping the tSZ component intact. We then investigate the dependence of CIB leakage in the resulting ILC $y$-map on the changes made in the frequency maps, which we can then use to infer $\beta$. As detailed later, we construct two ILC $y$-maps with $\beta$ deprojection\footnote{Here we use the terminology ``$\beta$ deprojection'' as shorthand for deprojecting the CIB with an MBB SED using value $\beta$.}: one from the original frequency maps, and one from the new set of frequency maps with altered CIB contributions. As we will show, if the $\beta$ value we have selected accurately cleans the CIB for the purpose of cross-correlating the resulting $y$-map with a given map of halos (or tracers), cross-correlating the difference in the two $y$-maps with the map of halos should give a result that is close to zero, since the CIB leakage in the two $y$-maps should be nearly identical. The correct $\beta$ for this cross-correlation is thus one that removes portions of the CIB that are correlated with the halos, thus yielding an unbiased measurement.

The overall process of determining the optimal CIB SED and constructing the resulting $y$-map consists of three main steps (each further detailed in the following subsections):
\begin{enumerate}
    \item Altering CIB Contributions: First, we devise a method to construct frequency maps with altered levels of CIB emission.  We then build $y$-maps both from these modified frequency maps and from the original frequency maps.
    \item Determining the Optimal $\beta$ Value: Next, using the two $y$-map variants, we discuss how to find the $\beta$ value in each multipole bin that maximally removes the CIB contributions to a given cross-correlation.
    \item Constructing the Final Map: Finally, we use the determined $\beta$ values to construct the final $y$-map, deprojecting a different CIB SED in each multipole bin.
\end{enumerate}

\subsection{Altering CIB Contributions}

Here we develop a data-driven approach for constructing frequency maps that have altered CIB contributions while keeping the tSZ contributions intact, i.e., the original frequency maps but with the CIB terms changed. This allows construction of two $y$-maps: one from the original frequency maps, and one from the frequency maps with altered CIB contributions. 

For simplicity, let us start by assuming there is some $\beta^*$ that is optimal (in the sense that it perfectly describes the CIB SED) across the entire frequency range and multipole range of interest. The temperature maps at each frequency are then given by
\begin{equation}
    T_\nu (\mathbf{\hat{n}}) = f_\nu y(\mathbf{\hat{n}}) + g^{\beta^*}_\nu c(\mathbf{\hat{n}}) + r_\nu(\mathbf{\hat{n}})+ n_\nu(\mathbf{\hat{n}}) \, ,
\end{equation}
where $f_\nu$ is the tSZ SED, $y(\mathbf{\hat{n}})$ is the Compton-$y$ field, $g_\nu^{\beta^*}$ is the CIB SED with the optimal $\beta$ value $\beta^*$, $c(\mathbf{\hat{n}})$ is the CIB field with the SED factored out, $r_\nu(\mathbf{\hat{n}})$ represents the residual CIB field that is not captured by the simple modified blackbody SED, and $n_\nu(\mathbf{\hat{n}})$ is the noise (and other fields that are not correlated with the halos at the two-point level on small scales, e.g., the CMB or kSZ field).

We build a Compton-$y$ ILC map constructed from the frequency maps $T_\nu(\mathbf{\hat{n}})$ (with deprojected CIB assuming emissivity index with a value of $\beta$), which can be represented as follows:
\begin{equation}
    \label{eq.ybeta}
    y^\beta(\mathbf{\hat{n}}) = y_{\rm true}(\mathbf{\hat{n}}) + c^\beta(\mathbf{\hat{n}}) + n^\beta(\mathbf{\hat{n}}) \, ,
\end{equation}
where $y^\beta(\mathbf{\hat{n}})$ is the ILC $y$-map produced when deprojecting the CIB with an SED using value $\beta$, $y_{\rm true}(\mathbf{\hat{n}})$ is the true Compton-$y$ field, $c^\beta(\mathbf{\hat{n}})$ is the leakage of the CIB into the map after being deprojected with an SED using value $\beta$ (this leakage includes contributions from both $g^{\beta^*}_\nu c(\mathbf{\hat{n}})$ and $r_\nu(\mathbf{\hat{n}})$), and $n^\beta(\mathbf{\hat{n}})$ is the leakage of the noise (and other fields not correlated with the halos) into the map after the CIB is deprojected with an SED using value $\beta$. Note that here we neglect the contribution from other sources, such as radio galaxies, to the frequency maps.

We then compute the residual at each frequency, defined as the original frequency maps with the tSZ contribution subtracted, where the tSZ contribution is estimated from the $y$-map in Eq.~\eqref{eq.ybeta}:
\begin{align}
    R_\nu^\beta(\mathbf{\hat{n}}) &= T_\nu(\mathbf{\hat{n}}) - f_\nu y^\beta(\mathbf{\hat{n}}) \nonumber
    \\&=  g^{\beta^*}_\nu c(\mathbf{\hat{n}}) + r_\nu(\mathbf{\hat{n}}) + n_\nu(\mathbf{\hat{n}}) - f_\nu c^\beta(\mathbf{\hat{n}}) - f_\nu n^\beta(\mathbf{\hat{n}}) 
\end{align}
This yields residual maps $R_\nu^\beta(\mathbf{\hat{n}})$ that have no tSZ contributions.

Next, we multiply the residual by a dimensionless SED $h_\nu$ (described later) and add that to the temperature maps to get a new set of temperature maps, $T_\nu '(\mathbf{\hat{n}})$:
\begin{align}
    \label{eq.t_nu_prime}
    T_\nu '(\mathbf{\hat{n}}) &= T_\nu(\mathbf{\hat{n}}) + h_\nu R^\beta_\nu(\mathbf{\hat{n}})  \\
    &= f_\nu y(\mathbf{\hat{n}}) + (1+ h_\nu)(g^{\beta^*}_\nu c(\mathbf{\hat{n}}) + r_\nu(\mathbf{\hat{n}}) + n_\nu(\mathbf{\hat{n}}))  \nonumber
    \\&\qquad-  h_\nu f_\nu (c^\beta(\mathbf{\hat{n}}) + n^\beta (\mathbf{\hat{n}}) ) \nonumber
\end{align}
This result is very close to what we want (i.e., the original frequency maps but with the CIB terms changed). However, some cancellation may arise from the $- h_\nu f_\nu (c^\beta(\mathbf{\hat{n}}) + n^\beta(\mathbf{\hat{n}}))$ term. The $h_\nu$ here is critical so that that component does not have SED $f_\nu$ (if it did, the ILC procedure would preserve it, and there would be some cancellation due to the negative sign). The factor of $h_\nu$ mitigates this problem, but does not completely solve it since it could still partially propagate into the ILC map. Nevertheless, we have freedom in choosing $h_\nu$, so we choose it such that $h_\nu f_\nu$ is orthogonal to $f_\nu$:
\begin{equation}
    \label{eq.orthogonality}
    \sum_\nu f_\nu^2 h_\nu = 0 \, ,
\end{equation}
where the right-hand side of the equation has implied units of $\mathrm{K}^2$.

For simplicity, we set $h_\nu=\alpha$ for $\nu \neq \nu_0$ (where $\nu_0$ is one arbitrarily selected frequency band and $\alpha$ is a nonzero real number), and 
\begin{equation}
    h_{\nu_0} = -\frac{1}{f_{\nu_0}^2}\sum_{\nu' \neq \nu_0} \alpha f_{\nu'}^2 \,.
\end{equation}
The choice to parametrize $h_\nu$ in this way is made for simplicity. In reality, there are infinite values that $h_\nu$ could take on, while still satisfying the orthogonality condition in Eq.~\eqref{eq.orthogonality}. One good approach would be to find $h_\nu$ that minimizes the variance of the ILC $y$-map built from frequency maps $T_\nu'$. This would be difficult to solve analytically while requiring $h_\nu \neq 0$, but could be solved numerically. We explore a more rigorous way to choose $h_\nu$ in Appendix \ref{app.h_nu}. The method described there leads to slightly smaller (by around 15\%) error bars on the optimal $\beta$ values, but in practice, we find that our simple parametrization  achieves unbiased $\beta$ values and nearly identical signal-to-noise in the final cross-correlation as compared to the more rigorous approach.

Next, we build an ILC $y$-map from $T_\nu'(\mathbf{\hat{n}})$, preserving the tSZ field and deprojecting $(1+ h_\nu) g^\beta_\nu$. We denote the resulting ILC map as $y^\beta_\alpha(\mathbf{\hat{n}})$:
\begin{equation}
    y^\beta_\alpha(\mathbf{\hat{n}}) = y_{\rm true}(\mathbf{\hat{n}}) + c^\beta_\alpha(\mathbf{\hat{n}}) + n^\beta_\alpha(\mathbf{\hat{n}}) \, ,
\end{equation}
where the interpretation of the terms is the same as in Eq.~\eqref{eq.ybeta}, but with leakage terms coming from the primed frequency maps with altered CIB component.

We perform this procedure for a range of physically reasonable $\beta$ values, constructing $y^\beta(\mathbf{\hat{n}})$ and $y^\beta_\alpha(\mathbf{\hat{n}})$ separately for each one.

\subsection{Determining the Optimal $\beta$ Values}

Having constructed both $y^\beta(\mathbf{\hat{n}})$ and $y^\beta_\alpha(\mathbf{\hat{n}})$ for a range of $\beta$ values, we then compute the difference between the two $y$-maps for each value of $\beta$: $y^\beta(\mathbf{\hat{n}})-y^\beta_\alpha(\mathbf{\hat{n}})=(c^\beta(\mathbf{\hat{n}}) - c^\beta_\alpha(\mathbf{\hat{n}})) + (n^\beta(\mathbf{\hat{n}}) - n^\beta_\alpha(\mathbf{\hat{n}}))$. Next we cross-correlate this difference with the halo or tracer map of interest. Here we use a map of halos $h(\mathbf{\hat{n}})$:
\begin{align}
    \label{eq.null_xcorr}
    C_\ell^{(y^\beta-y^\beta_\alpha),h} &= C_\ell^{(c^\beta-c^\beta_\alpha),h} + C_\ell^{(n^\beta-n^\beta_\alpha),h} 
\end{align}
We bin this power spectrum in multipole bins of width $\Delta \ell$. When $\beta =\beta^*$, the CIB leakage into each the two $y$-maps should have close to zero correlation with $h(\mathbf{\hat{n}})$, and thus, $C_\ell^{(c^\beta-c^\beta_\alpha),h}$ is very small. In the case that there is nonzero leakage $c^\beta(\mathbf{\hat{n}})$, the addition of the residual maps in Eq.~\eqref{eq.t_nu_prime} changes the effective SED of the CIB; thus, $c_\alpha^\beta(\mathbf{\hat{n}})$ is expected to be different from $c^\beta(\mathbf{\hat{n}})$, leading to nonzero $C_\ell^{(c^\beta-c^\beta_\alpha),h}$.

The first term in Eq.~\eqref{eq.null_xcorr} is dominant since the second term is zero on average, aside from fluctuations (note that here we have defined $n(\mathbf{\hat{n}})$ as only containing components that are not correlated with the halos, but we discuss including other components that are correlated with the halos in Sec.~\ref{sec.discussion}). These fluctuations grow as larger values of $\alpha$ are used, since $n^\beta_\alpha(\mathbf{\hat{n}})$ is much larger than $n^\beta(\mathbf{\hat{n}})$ for large $\alpha$ (this can be seen from Eq.~\eqref{eq.t_nu_prime}, where $h_\nu$, which is proportional to $\alpha$, inflates not only the CIB but also the noise terms). On the other hand, larger values of $\alpha$ correspond to larger differences between $c^\beta(\mathbf{\hat{n}})$ and $c^\beta_\alpha(\mathbf{\hat{n}})$, amplifying differences between the two maps.

We repeat this procedure for several values of $\beta$. For each $\beta$, we compute the $\chi^2$ of $C_\ell^{(y^\beta-y^\beta_\alpha),h}$ with respect to null, where the covariance matrix is computed using some fiducial $\beta'$. Specifically, 
\begin{align}
    \chi^2 &= C_\ell^{(y^\beta-y^\beta_\alpha),h} \nonumber
    \\&  \times \mathrm{Cov}^{-1} (C_\ell^{(y^{\beta'}-y^{\beta'}_\alpha),h}, C_\ell^{(y^{\beta'}-y^{\beta '}_\alpha),h} ) C_\ell^{(y^\beta-y^\beta_\alpha),h} \, .
\end{align}
We approximate the covariance matrix as Gaussian (and hence diagonal in $\ell$): 
\begin{align}
    \label{eq.cov}
    &\mathrm{Cov}(C_\ell^{(y^{\beta'}-y^{\beta'}_\alpha),h}, C_\ell^{(y^{\beta'}-y^{\beta '}_\alpha),h}) \nonumber \\ 
    &=\frac{1}{N_{\mathrm{modes}}} \left[ \left( C_\ell^{(y^{\beta'}-y^{\beta'}_\alpha),h} \right)^2 + C_\ell^{(y^{\beta'}-y^{\beta'}_\alpha),(y^{\beta'}-y^{\beta'}_\alpha)} C_\ell^{hh}  \right] \, ,
\end{align}
where $N_{\mathrm{modes}}=(2\ell+1)\Delta \ell$ and $\beta'$ is some fixed value that is used in the covariance matrix for all $\chi^2$ computations (i.e., the same covariance matrix is used to compute $\chi^2$ for every value of $\beta$). To be thorough, one could perform an iterative process for finding $\beta'$. First, pick an arbitrary $\beta'$ (within physical reason) and run the pipeline using that covariance matrix. Find $\beta^{*}$ that minimizes the $\chi^2$. Then rerun the pipeline with $\beta'=\beta^{*}$. Whether or not this iterative process is used, we find that the results are largely robust to the $\beta'$ used in the covariance calculation (and moreover, the covariance matrix is only needed for estimating the error bar on $\beta^*$ and not its central value, as will be clear below).

If there is no decorrelation of the CIB across frequencies and the CIB is perfectly described by an MBB with emissivity index $\beta^*$, then deprojecting the CIB with $\beta^*$ results in $c^\beta(\mathbf{\hat{n}})=c^\beta_\alpha(\mathbf{\hat{n}})=0$, and therefore $C_\ell^{(y^\beta-y^\beta_\alpha),h}=0$ aside from the noise fluctuations. This is \emph{only} true when $\beta=\beta^*$, and thus, it is clear in this case that the optimal value minimizes the $\chi^2$ of $C_\ell^{(y^\beta-y^\beta_\alpha),h}$ with respect to null. In this case of no CIB decorrelation, it does not matter what halo sample or tracer is used, since $c^{\beta^*}(\mathbf{\hat{n}})-c^{\beta^*}_\alpha(\mathbf{\hat{n}})=0$ already at the map level.

In the more realistic case that there is spatial variation of the CIB SED and/or decorrelation across frequencies, $c^\beta(\mathbf{\hat{n}})$ and $c^\beta_\alpha(\mathbf{\hat{n}})$ will have nonzero contributions from $r_\nu(\mathbf{\hat{n}})$, regardless of the optimality of the $\beta$ value chosen. Since $r_\nu(\mathbf{\hat{n}})$ is small relative to $g^{\beta^*}c(\mathbf{\hat{n}})$, $c^\beta(\mathbf{\hat{n}})$ and $c^\beta_\alpha(\mathbf{\hat{n}})$ both predominantly contain contributions from $g^{\beta^*}c(\mathbf{\hat{n}})$. However, since there is still no MBB SED that perfectly describes the CIB in this case, there is no $\beta$ such that $c^\beta(\mathbf{\hat{n}})-c^\beta_\alpha(\mathbf{\hat{n}})=0$ at the map level. Nevertheless, given the goal of optimizing $\beta$ for a specific cross-correlation, there \emph{is} some $\beta$ value, $\beta^*$, such that $c^{\beta^*}(\mathbf{\hat{n}})$ and $c^{\beta^*}_\alpha(\mathbf{\hat{n}})$ both have close to zero cross-correlation with $h(\mathbf{\hat{n}})$. When $\beta \neq \beta^*$, $c^{\beta}(\mathbf{\hat{n}})$ and $c^{\beta}_\alpha(\mathbf{\hat{n}})$ will each still have components that are correlated with $h(\mathbf{\hat{n}})$. Since $c^{\beta}(\mathbf{\hat{n}})$ and $c^{\beta}_\alpha(\mathbf{\hat{n}})$ are different by consequence of coming from frequency maps with different amounts of CIB, the further $\beta$ is from $\beta^*$, the more highly correlated $c^{\beta}(\mathbf{\hat{n}})-c^{\beta}_\alpha(\mathbf{\hat{n}})$ will be with the halos. Thus, $\beta=\beta^*$ still minimizes the $\chi^2$, though there may be a larger uncertainty on the optimal value of $\beta$ due to small contributions from $r_\nu(\mathbf{\hat{n}})$.

Once the maps $y^\beta(\mathbf{\hat{n}})$ and $y^\beta_\alpha(\mathbf{\hat{n}})$ have been produced, the process of determining the optimal value of $\beta$ (by finding $\beta$ that minimizes the $\chi^2$ as described above) is performed separately in each multipole bin. This is necessary because in realistic CIB maps, the effective value of $\beta$ may change as a function of $\ell$. Indeed, we find this to be true in simulations as shown in Sec.~\ref{sec.results}. We denote the optimal $\beta$ in each multipole bin as $\beta^*_\ell$.

We can also assess the uncertainty on the $\beta^*_\ell$ by finding the $\beta$ values with $\Delta \chi^2=1$ as compared with the minimum $\chi^2$, which is attained at $\chi^2(\beta^*_\ell)$. This gives the $1\sigma$ uncertainty on $\beta^*_\ell$ in each multipole bin.

\subsection{Constructing the Final Map}

To construct the final $y$-map, we perform harmonic ILC with one deprojection of the CIB, a contaminant with an $\ell$-dependent SED. At each $\ell$, we deproject $\beta^*_\ell$ that was obtained for the bin in which the multipole belongs. Thus, all multipoles in a bin have the same SED deprojected. 

It is straightforward to generalize the harmonic ILC procedure for deprojecting such a contaminant in an $\ell$-dependent way. Specifically, Eq.~\eqref{eq.cilc} becomes
\begin{widetext}
\begin{equation}
\label{eq.cilc_ell_dep}
    w^j_{\ell} = \frac{ \left((g_\ell)_k (\hat{R}_{\ell}^{-1})_{km} (g_\ell)_m \right) f_i (\hat{R}_{\ell}^{-1})_{ij} - \left(f_k (\hat{R}_{\ell}^{-1})_{km} (g_\ell)_m \right) (g_\ell)_i (\hat{R}_{\ell}^{-1})_{ij}}{\left(f_k (\hat{R}_{\ell}^{-1})_{ki} f_i\right) \left((g_\ell)_m (\hat{R}_{\ell}^{-1})_{mn} (g_\ell)_n \right) - \left(f_k (\hat{R}_{\ell}^{-1})_{ki} (g_\ell)_i \right)^2} \, ,
\end{equation}
\end{widetext}
where the difference is just that the contaminant SED $g_i$ becomes an $\ell$-dependent quantity, $(g_\ell)_i$. In particular,  
\begin{equation}
   (g_\ell)_i  = \left( \frac{\nu_i}{\nu_0} \right)^{\beta^*_\ell+3} \frac{1}{e^{h\nu_i/(k_B T_{d})}-1} \, .
\end{equation}

For the purpose of this work, in each bin we deproject the central value of $\beta^*_\ell$ without considering the error bar. In theory it is also possible to perform a smoothing operation or to fit some function for $\beta$ as a function of $\ell$ so that the deprojection is less sensitive to the noise in each bin. However, it is not clear what functional form to use, or what smoothing operation to perform. We thus leave this idea to future work, and here only consider using the central values of $\beta^*_\ell$ with sufficiently wide multipole bins to mitigate the effects of noise.  This approach is also motivated by the fact that $\beta^*_\ell$ is a slowly varying function of $\ell$.

For notational simplicity, from here on we refer to this final $y$-map as $y^{\beta^*}$, even though it is really a map with $\beta^*_\ell$ deprojected separately for each bin. We can then cross-correlate $y^{\beta^*}$ with $h$, and estimate the error bars on the cross-correlation using the same Gaussian approximation as in Eq.~\eqref{eq.cov}. 

It is critical to note that this procedure finds $\beta^*_\ell$ that minimizes CIB contamination in the specific cross-correlation with $h$, in particular, by selecting $\beta^*_\ell$ that maximally removes portions of the CIB that are correlated with $h$. It does \emph{not} find any ``true" global $\beta^*_\ell$. Thus, the final $y$-map here is \emph{only} valid for cross-correlating with $h$, and should not be thought of as an optimal $y$-map itself nor used in any other cross-correlation. If used in a cross-correlation with a different halo selection or different tracer, there would still be residual CIB in the map that is correlated with the new halo selection, since the procedure has only cleaned contributions of the CIB that are correlated with the original halo selection, $h$.

\section{Simulation Set-Up}

\label{sec.simulations}

\begin{figure*}[htb]
    \centering
    \includegraphics[width=1.0\textwidth]{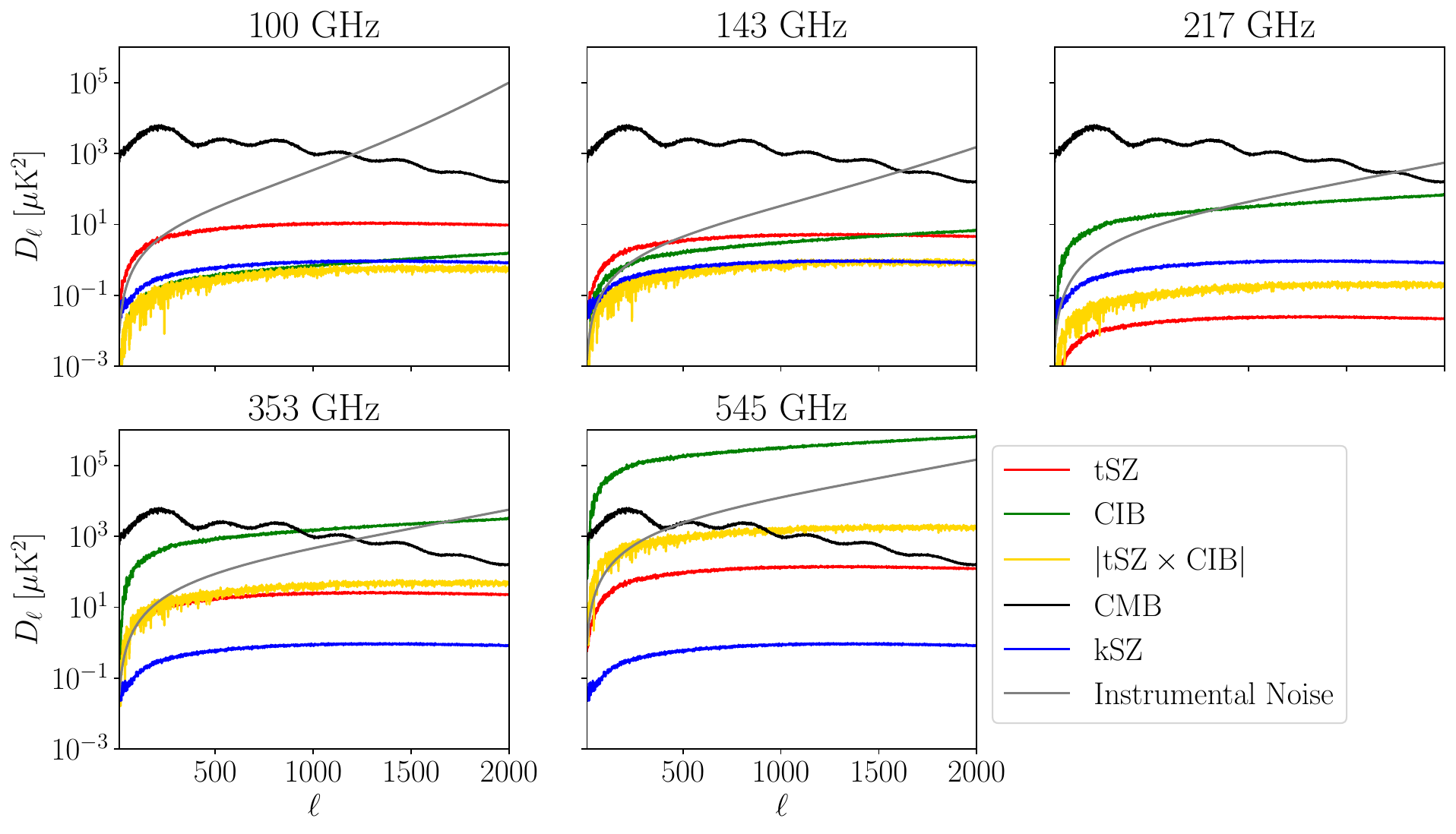}
    \caption{Power spectra (plotted as $D_\ell = \ell(\ell+1)C_\ell/(2\pi) $ in units of $\mu \mathrm{K}^2$) of simulation components (with \emph{Planck} passbands, but without multipole binning) at the frequency channels used in this work. The components shown are the tSZ effect (red), CIB (green), $|\mathrm{tSZ} \times \mathrm{CIB}|$ (gold), CMB (black), kSZ effect (blue), and instrumental noise (gray). The CIB is the dominant component at high frequencies. The extragalactic foregrounds are from the \texttt{AGORA} simulation suite \cite{Omori:2022}.}
    \label{fig:comp_ps}
\end{figure*}

We test our algorithm on simulations containing the tSZ effect, CIB, kSZ effect, CMB, and \emph{Planck} instrumental noise. We use simulated maps of the five \emph{Planck} high-frequency channels: 100, 143, 217, 353, and 545 GHz. The maps used have \texttt{healpix} \cite{Healpy, Healpix} resolution parameter $N_{\rm side}=1024$. For simplicity, none of the simulated maps are beam-convolved, and multipoles $2 \leq \ell \leq 2000$ are used in this analysis. We use component maps from the \texttt{AGORA} simulation suite \cite{Omori:2022}. The power spectra of the component maps are shown in Fig.~\ref{fig:comp_ps}. 

For the \texttt{AGORA} simulations, halo catalogs are generated from the \texttt{Multidark-Planck2} (MDPL2) $N$-body simulation, which uses a box size of 1 $h^{-1}$ Gpc, $3840^3$ particles, and mass resolution of $1.51 \times 10^9 \,\, h^{-1} M_{\odot}$ \cite{Klypin:2014kpa}.

The tSZ map is generated by using halos with masses $M > 10^{12} \,\, h^{-1}M_{\odot}$ and pasting electron pressure profiles onto those halos. We use a variant of the model from Ref.~\cite{Mead:2020} with AGN heating temperature $10^{8.0}$~K in the \texttt{BAHAMAS} simulation \cite{bahamas}. It is simulated at redshifts $0 < z < 3.0$.

For the CIB, the \texttt{UniverseMachine} code is used to assign a star formation rate to each halo \cite{universemachine}, and then the star formation rates are converted into infrared luminosities \cite{Kennicutt:1997ng}. The CIB is simulated at redshifts $ 0 < z < 8.6$. The SED of each individual infrared galaxy is modeled as a modified blackbody with a power-law transition as in \cite{2016A&A...594A..15P}:
\begin{align}
g_\nu =  \begin{cases}
  \left[\exp{(\frac{h\nu}{k_BT_d})-1} \right]^{-1} \nu^{\beta+3} & \nu \leq \nu' \\
   \left[\exp{(\frac{h\nu'}{k_BT_d})-1} \right]^{-1} \nu^{'\beta+3} (\frac{\nu}{\nu'})^{-\alpha_d} & \nu > \nu' \, ,
\end{cases}
\end{align}
where $\nu'$ is the frequency at which the power-law transition occurs. In the model, $\beta$ and $T_d$ are assumed to be correlated, similarly to \cite{Dupac:2003ek}:
\begin{equation}
    \beta = \frac{\zeta_d}{0.4 + 0.008T_d} \, .
\end{equation}
Then $\alpha_d$, $\zeta_d$, and other free parameters of the CIB model are fit to the auto- and cross-spectra at frequencies 353, 545, and 857 GHz from \cite{Lenz:2019ugy}.

\begin{figure}[t]
    \centering
    \includegraphics[width=0.48\textwidth]{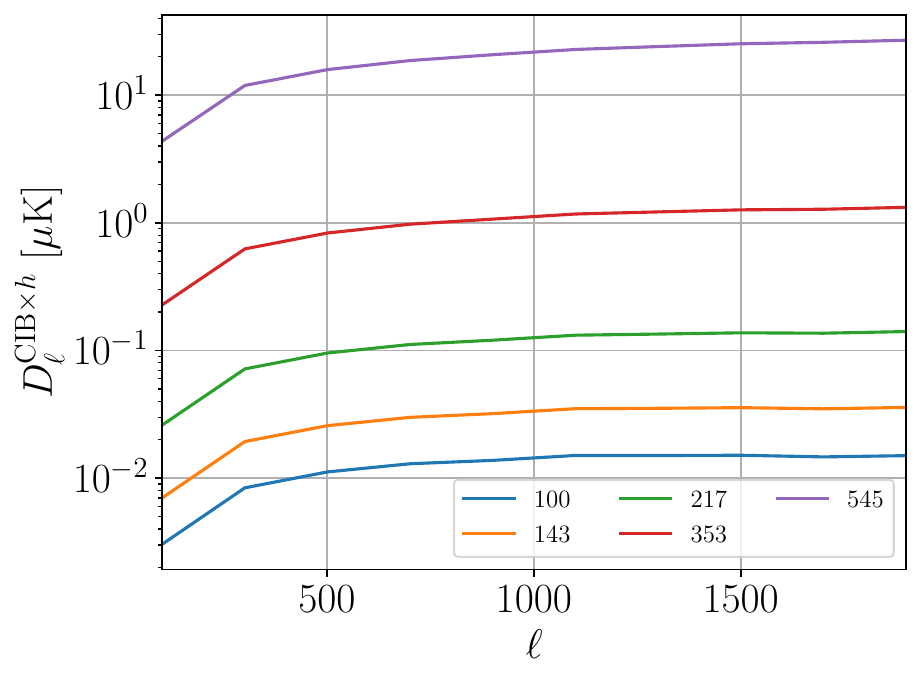}
    \caption{Cross-power spectrum of the CIB with halos (plotted as $D_\ell = \ell(\ell+1)C_\ell/(2\pi) $ in units of $\mu \mathrm{K}$) for each of the frequency channels used in this work. The halos selected have $0.8 < z < 1.8$ and $10^{12} M_{\odot} < M < 10^{15} M_{\odot}$.}
    \label{fig:cibxh}
\end{figure}

For the baseline halo density map with which to cross-correlate the ILC maps, we use the \texttt{AGORA} shells 1 through 120, selecting halos in the redshift range $0.8 < z < 1.8$ and mass range $10^{12} M_{\odot} < M < 10^{15} M_{\odot}$, a total of 279,851,713 halos. This allows for the cross-correlation of the CIB and halos to be positive up to high $\ell$ for all the frequency channels we use, as shown in Fig.~\ref{fig:cibxh}.\footnote{For lower redshift halos, the \texttt{AGORA} simulation cross-correlation of CIB and halos becomes negative at much lower values of $\ell$. However, as shown in Appendix~\ref{app:diff_halos}, our procedure still works as long as we restrict our analysis to multipoles such that the cross-correlation is positive, albeit at the cost of losing some signal-to-noise. We note that the contaminant--tracer cross-correlation does not strictly need to be positive, but that it does need to have the same sign across frequencies to avoid cancellation that may arise when summing over frequencies in the ILC procedure.} In Appendix \ref{app:diff_halos} we show that the procedure works on lower-redshift halo selections as well.

For the ILC procedure, we use harmonic ILC for multipoles $2 \leq \ell \leq 2000$ in bins of width $\Delta \ell \approx 200$. The maximum multipole $\ell = 2000$ is chosen to be approximately two times the $N_{\rm side}$ of the maps. We assume that the CIB takes on the SED of an MBB of temperature 24.0 K.\footnote{The assumption that $T_d=24.0$ K implies that the optimal $\beta$ we find is the optimal $\beta$ for that temperature. There is a significant degeneracy between the two parameters over the frequency range studied here, so changing $T_d$ would change the optimal $\beta$ value obtained. Thus, the $\beta$ value we find should not be thought of as a ``true" physically meaningful parameter, but rather, as a best-fit for the specific cross-correlation, given the parametrization and assumed temperature.} For all SED calculations, we use the \emph{Planck} passbands \cite{planck_2013_ix} since those were used in construction of the simulated maps.

We test the procedure on 84 values of $\beta$ between 0 and 3, most densely sampled in the region $1.5 \leq \beta \leq 1.9$ (with a spacing of 0.0125 between each tested $\beta$ value in that region), and less densely sampled in the regions $0 \leq \beta \leq 1.5$ and $1.9 \leq \beta \leq 3.0$ (with a spacing of 0.05 between each tested $\beta$ value in that region). For each $\beta$, we construct $y^\beta$ and $y^\beta_\alpha$, compute $C_\ell^{(y^\beta- y^\beta_\alpha),h}$, and finally compute the $\chi^2$ between $C_\ell^{(y^\beta-y^\beta_\alpha),h}$ and a vector of zeros. We use $\beta'=1.70$ for computation of the covariance matrix. 

For $\nu_0$, defined in Sec.~\ref{sec.algorithm}, we select the 545 GHz channel, and for $\alpha$ (also defined in Sec.~\ref{sec.algorithm}), we demonstrate the algorithm on $\alpha \in \{ 0.1, 1.0, 10.0 \}$, showing that it is largely insensitive to the $\alpha$ value used (we discuss alternate parametrizations of $h_\nu$ in Appendix \ref{app.h_nu}). We find $\beta^*_\ell$ that minimizes the $\chi^2$ in each bin via cubic interpolation of the $\chi^2$ curve over the range of $\beta$ values tested. To find the $1\sigma$ range, we find the $\beta$ values that have $\Delta \chi^2=1$ with respect to $\chi^2(\beta^*_\ell)$.

To assess the validity of these results, we run an idealized test. First, we construct frequency maps that contain all components except the CIB, and build a standard ILC $y$-map (no deprojection is needed since there is no CIB) from those frequency maps. We denote this $y$-map as $y^{\rm opt}$. This map is not possible to construct from actual data, but from a theoretical standpoint it is useful for comparison here since we know there is no CIB in the final map, i.e., the final map will only contain the true Compton-$y$ field plus some leakage from the CMB, kSZ effect, and instrumental noise. It is thus a theoretically best ILC $y$-map one can build given the noise. For every value of $\beta$ tested in the primary pipeline, we compute $C_\ell^{(y^\beta-y^{\rm opt},h)}$, and then compute the $\chi^2$ between that and a vector of zeros. The covariance matrix here is 
\begin{align}
    &\mathrm{Cov}(C_\ell^{(y^{\beta'}-y^{\rm opt}),h}, C_\ell^{(y^{\beta'}-y^{\rm opt}),h}) \nonumber \\ 
    &\qquad = \frac{1}{N_{\mathrm{modes}}} \Bigg[ \left( C_\ell^{(y^{\beta'}-y^{\rm opt}),h} \right)^2  \nonumber \\
    &\qquad \qquad \qquad \qquad  + C_\ell^{(y^{\beta'}-y^{\rm opt}),(y^{\beta'}-y^{\rm opt})} C_\ell^{hh}  \Bigg] \, ,
\end{align}
where $\beta'=1.70$ just as in the main pipeline. We then find $\beta^{*}_\ell$ and the $1\sigma$ uncertainty in the same way as in the main pipeline. We again emphasize this is an extremely idealized test and can never be performed on actual data. It is performed here solely for validation of the primary algorithm, where we demonstrate that $\beta^{*}_\ell$ from the two approaches agree reasonably well.

\section{Results}
\label{sec.results}

\begin{figure*}[t]
    \centering
    \includegraphics[width=0.95\textwidth]{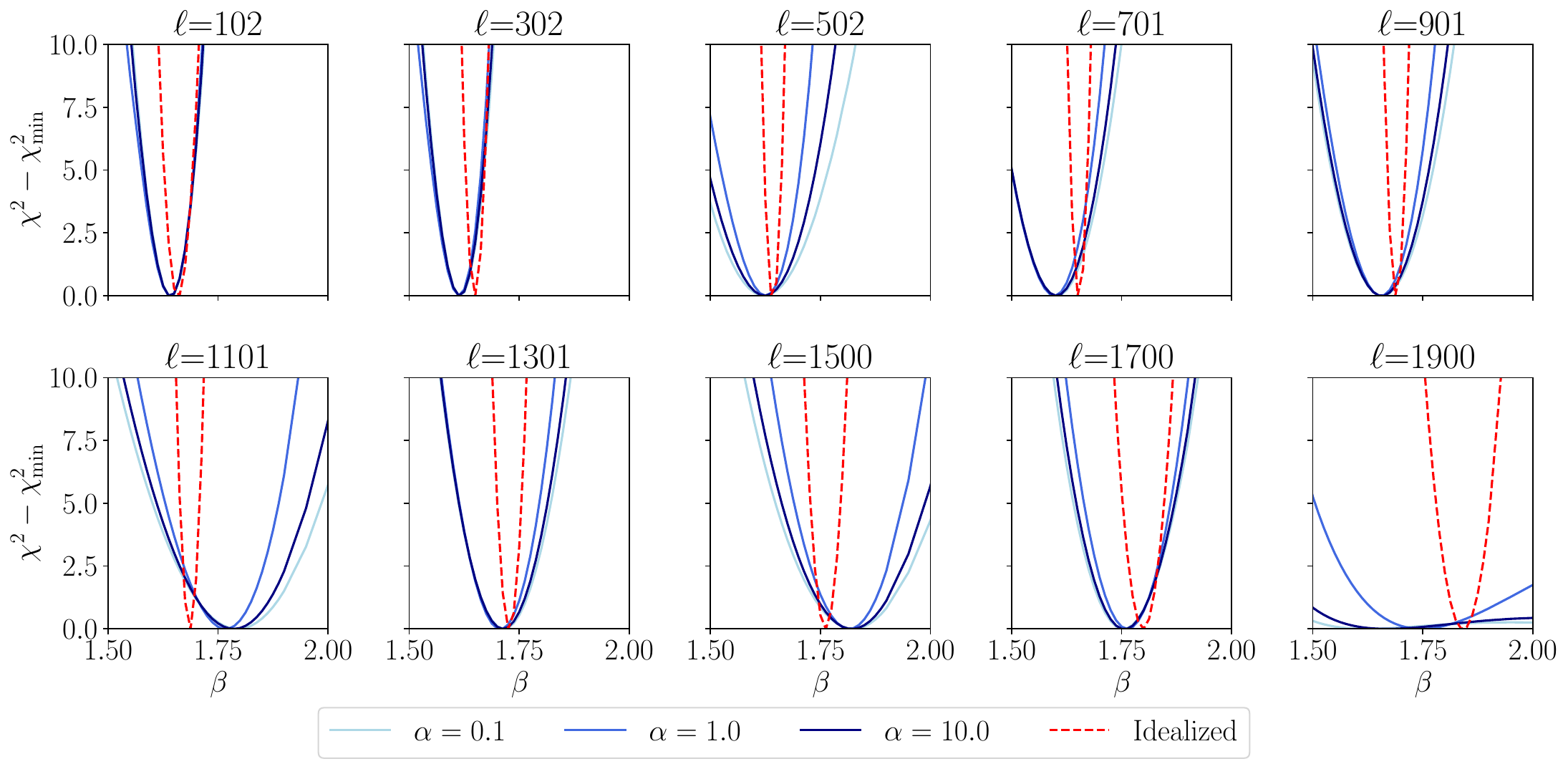}
    \caption{$\chi^2$ calculations for each value of $\beta$ used in the ILC CIB deprojection procedure, with the minimum $\chi^2$ of each curve subtracted such that the minimum of each curve is zero. The ILC maps are built from frequency maps containing the tSZ effect, CIB, kSZ effect, CMB, and \emph{Planck} HFI instrumental noise. The procedure is performed separately in each multipole bin, with the subplot titles denoting the mean multipole in each bin. The red curves are from an idealized procedure, comparing $y^{\beta}$ (the ILC $y$-map with $\beta$ deprojection, constructed from the full frequency maps) to $y^{\rm opt}$ (the standard ILC $y$ map constructed from idealized frequency maps that have no CIB). The blue curves compare $y^{\beta}_\alpha$ (the ILC $y$-maps with $\beta$ deprojection, constructed from frequency maps with inflated residual) to $y^{\beta}$. These curves are shown for various values of $\alpha$, the parameter that sets the inflation of the residual (see Sec.~\ref{sec.algorithm}).}
    \label{fig:chi2_vs_beta}
\end{figure*}

\begin{figure}[htb]
    \centering
    \includegraphics[width=0.48\textwidth]{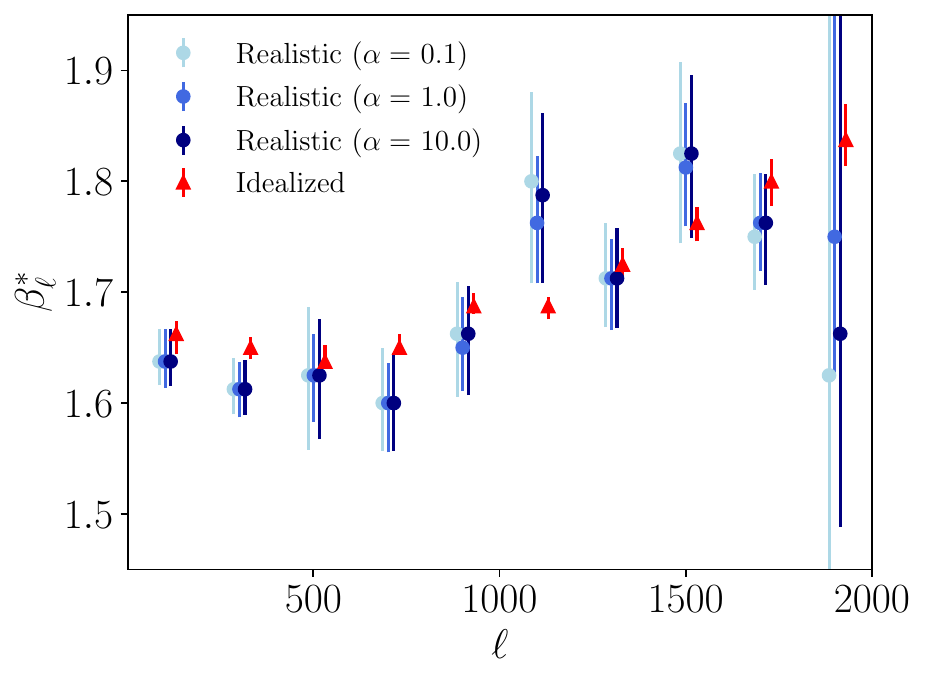}
    \caption{$\beta^{*}_\ell$ and its $1\sigma$ uncertainty range for various values of $\alpha$ (shades of blue) in each multipole bin. The red points and error bars show the $1\sigma$ uncertainty range of $\beta^{*}_\ell$ determined from an idealized scenario that cannot be applied to actual data, but is shown here for comparison. The points have a slight horizontal offset from one another for visual clarity.}
    \label{fig:beta_vs_ell}
\end{figure}

\begin{figure*}[htb]
    \centering
    \includegraphics[width=1.0\textwidth]{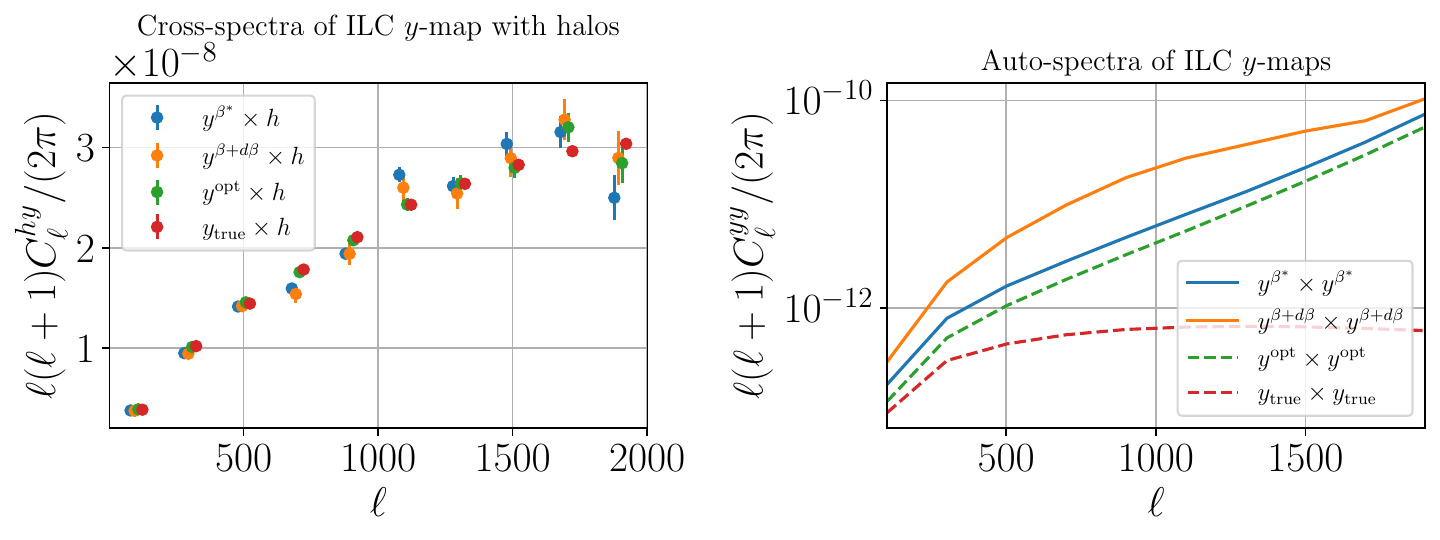}
    \caption{Comparison of our new method with ILC techniques that deproject both $\beta$ and $d\beta$. The map constructed from our new algorithm is denoted $y^{\beta^*}$ (blue). The map with moment deprojection is denoted $y^{\beta + d\beta}$ (orange). The standard ILC map built from frequency maps with noise but without CIB is denoted $y^{\rm opt}$ (green). The true $y$-map is denoted $y_{\rm true}$ (red). We note that only $y^{\beta^*}$ (blue) and $y^{\beta + d\beta}$ (orange) can actually be constructed from data. The other points are shown just for comparison. \textbf{Left:} Cross-spectra of ILC $y$-maps with halos, with Gaussian error bars. The differently colored points have slight horizontal offsets for visual clarity. \textbf{Right:} Auto-spectra of ILC $y$-maps (including noise and residual foreground contributions, where present).}
    \label{fig:dbeta}
\end{figure*}

\begin{figure}[htb]
    \centering
    \includegraphics[width=0.4\textwidth]{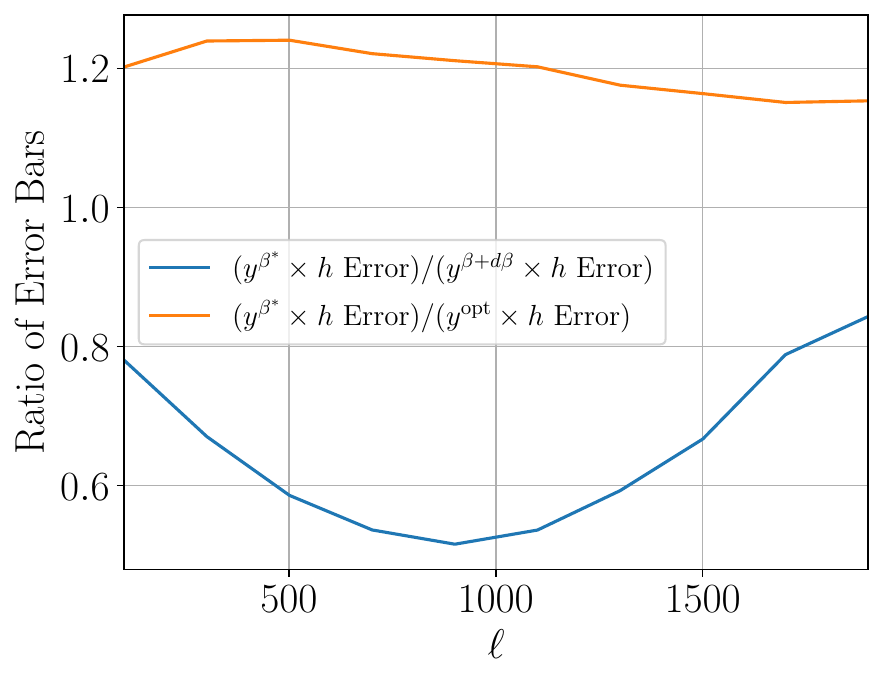}
    \caption{Comparison of error bars for $y^{\beta^*} \times h$, $y^{\beta + d\beta} \times h$, and $y^{\rm opt} \times h$. Using $y^{\beta^*}$ instead of $y^{\beta + d\beta}$ leads to significantly smaller error bars across the whole multipole range. We note that $y^{\rm opt}$ is the theoretical best error bar one could achieve, but is not possible to construct with actual data. }
    \label{fig:error_ratio}
\end{figure}

Fig.~\ref{fig:chi2_vs_beta} shows how $\chi^2$ varies as a function of $\beta$ for various values of $\alpha$, as well as $\chi^2$ as a function of $\beta$ for the idealized pipeline. Prior to subtracting $\chi^2_{\rm min}$ for each curve, the idealized curve has notably higher $\chi^2$ values than those from the main pipeline. This makes sense because the idealized curve is comparing a standard ILC map to a constrained ILC map. While, by construction, both maps should have no CIB (or close to no CIB in the constrained ILC maps when $\beta_\ell$ is sufficiently close to $\beta^{*}_\ell$), the constrained ILC procedure significantly increases the noise in the resulting map. Thus, the difference in the two maps has a large amount of noise, whereas much of this noise is canceled in the difference $y^\beta-y^\beta_\alpha$ since both involve constrained ILC procedures.

Fig.~\ref{fig:beta_vs_ell} shows the values of $\beta^{*}_\ell$ and $1\sigma$ uncertainty ranges obtained from pipelines using various values of $\alpha$. In all cases, the results agree with the idealized case (in red) within just over $1\sigma$. Importantly, we note that, regardless of $\alpha$, the inferred $\beta^{*}_\ell$ remains unbiased. The difference is primarily in the error bars, which vary as a function of $\alpha$, though not by much. For the remainder of this section, we focus only on the $\alpha=1$ case since the change in results is negligible with different values of $\alpha$.

We find that, using this new algorithm, tSZ -- halo cross-correlations achieve significantly higher signal-to-noise than the standard method of deprojecting both $\beta$ and the first moment with respect to $\beta$, $d \beta$. To build the map with the latter, we use \texttt{pyilc} \cite{McCarthy:2023cwg, McCarthy:2023hpa}. Fig.~\ref{fig:dbeta} shows various metrics by which to compare this method with $d \beta$ deprojection. In the moment deprojection case, we use $\beta=1.70$, though the final map should be insensitive to the exact value used since the first moment is also deprojected. In the map constructed with the new algorithm, we deproject the best-fit $\beta^*_\ell$ from Fig.~\ref{fig:beta_vs_ell} in each multipole bin separately.

We find that our new algorithm yields a $y$-map with an unbiased cross-correlation with the halos, while having a significantly lower auto-spectrum and hence smaller error bars on the cross-correlation, due to the reduced noise in the map. To validate the unbiased nature of our method, we fit a simple model to the $y^{\beta^*} \times h$ points, which is comprised of an overall amplitude $A$ that multiplies the $y_{\rm true} \times h$ points. If the measurement is unbiased, $A$ should be consistent with unity. We find $A = 0.973 \pm 0.010$, thus confirming that our method recovers an unbiased cross-power spectrum, without the need to deproject the $d\beta$ component. With the moment deprojection approach, the bias is somewhat smaller due to the larger error bars ($-2.1\sigma$ as compared with $-2.7\sigma$ from the new approach). There is thus a bias-variance tradeoff between the two approaches. Nevertheless, we note that removing the point around $\ell=700$ significantly reduces the bias to $-1.0\sigma$ with the new method (and $-1.1\sigma$ in the moment deprojection case). Thus, the overall mild deviation from unity is driven by this single outlier point. In any case, our new method recovers the expected cross-correlation within $3\sigma$, making it an unbiased measurement with significantly higher signal-to-noise than the moment deprojection approach.

While not shown in Fig.~\ref{fig:dbeta}, it is important to note that the cross-power spectrum of $y^{\beta^*}$ with $y_{\rm true}$ (the true $y$-map used in the simulations) is biased, which is to be expected. Our method does not determine a ``true" underlying value of $\beta$ or ``true" $y$-map, but rather, is only optimized for cross-correlation with a specific tracer.  If the tracer sample comprised all of the halos that contribute to the total $y$ field, then the method could potentially be used to obtain an unbiased auto-spectrum, but this is likely beyond the scope of any upcoming LSS surveys.

Fig.~\ref{fig:error_ratio} compares the cross-correlation error bars obtained using $y^{\beta^*}$, $y^{\beta + d\beta}$, and $y^{\rm opt}$. Across the entire multipole range, $y^{\beta^*} \times h$ has significantly smaller error bars than $y^{\beta + d\beta} \times h$, ranging from approximately 50\% to 80\% the size.  We also show $y^{\rm opt} \times h$, since it is a theoretical limit on the best error bar one could achieve, though it is impossible to construct $y^{\rm opt}$ in practice. We find that the error bar on $y^{\beta^*} \times h$ is approximately 1.2 times that of $y^{\rm opt} \times h$. In terms of signal-to-noise, $y^{\beta^*} \times h$, $y^{\beta + d\beta} \times h$, and $y^{\rm opt} \times h$ have a signal-to-noise of 95, 58, and 118, respectively.  Thus, our new method yields an improvement of over 60\% in the total detection significance of the tSZ -- halo cross-correlation, compared to the approach of deprojecting $\beta + d\beta$.

\begin{figure*}[htb]
    \centering    \includegraphics[width=0.85\textwidth]{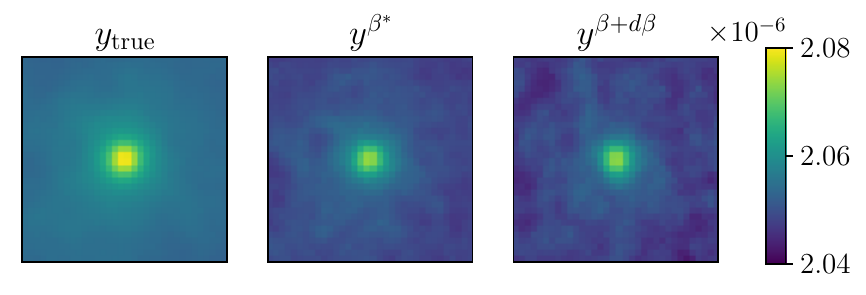}
    \caption{$y$-maps stacked on locations of halos, using a random sample of $\approx 2.8 \times 10^6$ halos (1\% of the halos) for computational efficiency. For each of the randomly selected halos, a 2D patch of 2 degrees on each side is cut out of the $y$-map around the location of the halo, and all the patches are averaged together. The left, center, and right images show the results of the stacking procedure on the true $y$-map ($y_{\rm true}$), the optimal $y$-map from our new method ($y^{\beta^*}$), and the $y$-map deprojecting both the CIB and its first moment with respect to $\beta$ ($y^{\beta + d\beta}$), respectively. Each of the $y$-maps are smoothed with a beam of FWHM 10 arcmin prior to performing the procedure.}
    \label{fig:stacking}
\end{figure*}

Fig.~\ref{fig:stacking} shows the results of stacking the different $y$-maps on the locations of the halos. In particular, we first smooth each of the maps with a Gaussian beam of FWHM 10 arcmin. We then randomly sample $2.8 \times 10^6$ (1\%) of the total halos, and cut 2 degree by 2 degree patches of the $y$-maps around the locations of the halos. These patches are stacked to produce the final images. We note that the background noise is significantly lower in $y^{\beta^{*}}$, the $y$-map produced deprojecting only the CIB with $\beta^{*}_\ell$ found via our algorithm, as compared with $y^{\beta + d\beta}$, the $y$-map produced by deprojecting both the CIB and its first moment $d\beta$.

In Appendix \ref{app:diff_halos}, we show results using a lower-redshift set of halos than what has been used in the baseline results here. We thus show that the utility of this algorithm is not restricted to the specific halo selection used here. In Appendix \ref{app:no_decorr}, we test the pipeline on simulations without CIB decorrelation, allowing us to know the true value of $\beta^*$ (which exactly exists in this case, since we model the CIB as a perfect MBB without decorrelation across frequencies). We show that the algorithm recovers the known true $\beta^*$, thus showing that it is unbiased.

\section{Conclusion}

\label{sec.discussion}

In this paper, we have devised an algorithm for determining CIB SEDs for deprojection in ILC $y$-maps, for use in cross-correlations. Specifically, we are able to determine values of $\beta$ in each multipole bin that are optimized specifically for obtaining unbiased, high signal-to-noise cross-correlations with a specific tracer. We have tested the algorithm on simulations, showing that the technique produces unbiased cross-correlations, with extensive robustness tests shown in the appendices. We find that the error bars in the cross-correlations from $y$-maps built with this new technique are 20-50\% smaller than those involving $y$-maps constructed with standard moment deprojection techniques, resulting in significant improvements in signal-to-noise (60\% in the example studied here).  We have used a \emph{Planck}-like experimental setup in this work, but we expect similar gains to be found for ground-based multifrequency CMB experiments like ACT \cite{dr6_maps, dr6_lcdm, dr6_extended}, SPT \cite{SPT-3G:2022hvq}, the Simons Observatory (SO) \cite{SimonsObservatory:2025wwn}, and CMB-S4 \cite{CMB_S4_Science_Book}.

In the algorithm described in Sec.~\ref{sec.algorithm}, we have assumed that the only component of the residual that is correlated with halos is the CIB. In reality, components such as radio sources~\cite{Li:2021ial, Omori:2022} and extragalactic CO lines violate this assumption \cite{Maniyar_CO, Kokron:2024}. One could still imagine performing a similar procedure, but noting that the effective $\beta$ found would contribute to the partial deprojection of all components correlated with halos. We leave this investigation to future work.

Another next step is generalizing the algorithm to needlet ILC (NILC) \cite{Delabrouille2009} procedures. Here we have found the optimal $\beta$ value to deproject in each multipole bin. A similar procedure could be applied in the NILC case, deprojecting a separate SED at each needlet filter scale. 

This method could then be applied to actual data, allowing high signal-to-noise measurements of several types of cross-correlations. In particular, when constructing $y$-maps from ground-based CMB experiments like ACT, SPT, or SO, there is high-resolution data on small scales, which is not present in satellite experiments like \emph{Planck}. The limitation in using the small-scale information for building $y$-maps is limited frequency coverage, since the ACT Data Release 6 (DR6), for example, used only three frequency bands \cite{dr6_maps}. To build a $y$-map with moment deprojection, at least three frequency channels are needed to satisfy the three constraints in the ILC procedure, and more are needed for meaningful signal-to-noise in the maps. Removing the moment deprojection constraint allows us to use one fewer frequency channel, thereby taking better advantage of the small-scale information in these experiments. 

Our method can potentially be extended beyond tSZ cross-correlations. It could be applied for tSZ auto-spectrum measurements if one has access to a halo sample that fully covers the range of halos that source the Compton-$y$ signal (as before, with the caveat that other signals correlated with the tSZ effect would also need to be cleaned; in addition, all other contributions to the auto-spectrum, such as from Galactic foregrounds and uncorrelated extragalactic foregrounds, also require treatment). The method could also be applied to building higher-SNR ILC maps for other cross-correlations, such as those searching for patchy screening or patchy dark screening of the CMB to place constraints on axions~\cite{Goldstein:2024mfp_axion, Mondino:2024rif} or dark photons~\cite{2024JCAP...01..019P, McCarthy:2024ozh_dark_photon}. Moreover, it could be used for CIB cleaning in projected-field kSZ estimators \cite{Dore:2003ex, DeDeo:2005yr, Hill:2016dta, Ferraro:2016ymw, Kusiak_2021, Patki:2023fjz}. Additionally, if one had a Galactic field that well traced the sources of polarized thermal dust emission in the Milky Way (e.g., HI maps \cite{Clark:2015cpa, Halal:2023sva} or stellar polarization maps \cite{Tassis2018, Pelgrims2024}), our technique could be used to infer the $\beta$ value of the polarized dust associated with these sources, by cross-correlating an ILC CMB $B$-mode map with that tracer. The resulting optimal $\beta$ value could then be used in CMB $B$-mode inference, thus negating the need to deproject multiple dust SED moments (e.g.,~\cite{2022A&A...660A.111V,2021JCAP...05..047A,2023JCAP...03..035A}), and hence tightening constraints on the tensor-to-scalar ratio $r$.  We leave such investigations to future work.

\section{Acknowledgments}

SP and JCH dedicate this work to the memory of EJB.

KS is supported by the National Science Foundation Graduate Research Fellowship Program under Grant No. DGE 2036197. JCH acknowledges support from NSF grant AST-2108536, NASA grants 21-ATP21-0129 and 22-ADAP22-0145, the Sloan Foundation, and the Simons Foundation.  The authors acknowledge the Texas Advanced Computing Center (TACC) at The University of Texas at Austin for providing HPC resources that have contributed to the research results reported within this paper. Several software tools were used in the development and presentation of results shown in this paper, including \verb|HEALPix/healpy| \cite{Healpix, Healpy}, \verb|numpy| \cite{numpy}, \verb|scipy| \cite{scipy}, \verb|matplotlib| \cite{matplotlib}, and \verb|astropy| \cite{astropy1, astropy2, astropy3}.


\onecolumngrid
\appendix

\section{Determining the Optimal $h_\nu$ Values}
\label{app.h_nu}
Here we discuss how to decide which $h_\nu$, defined in Eq.~\eqref{eq.t_nu_prime}, to use for adding the residual to the original frequency maps to obtain the primed frequency maps. As previously mentioned, one key constraint is the orthogonality of $f_\nu$ and $h_\nu f_\nu$, as in Eq.~\eqref{eq.orthogonality}. However, a formally infinite set of possible $h_\nu$ satisfy this constraint. We thus consider how to choose $h_\nu$ in an optimal way.

Let us consider some fiducial $\beta'$. One criterion for choosing $h_\nu$ is to choose it such that it minimizes the variance of the resulting ILC $y$-map, $y^{\beta'}_\alpha$. The variance of the $y$-map is given by
\begin{align}
    \mathrm{Var}(y^{\beta'}_\alpha) &= w^i_\ell w^j_\ell C_\ell^{ij} = \frac{d_k (\hat{R}^{-1}_\ell)_{ki}d_i}{(f_k (\hat{R}^{-1}_\ell)_{ki}f_i) (d_m (\hat{R}^{-1}_\ell)_{mn} d_n) - (f_k (\hat{R}^{-1}_\ell)_{ki}d_i)^2} \, 
\end{align}
where $d_i = (1+h_i)g^{\beta'}_i$. Minimizing this variance would yield an $\ell$-dependent SED $(h_\nu)_\ell$. While this is possible in theory, it is likely that such an $\ell$-dependence would lead to noise in the selection of $h_\nu$ if the multipole bin widths are not sufficiently large. Thus, here we find a single $h_\nu$ for all $\ell$ by performing a weighted average over multipoles. Rather than using $(\hat{R}^{-1}_\ell)_{ij}$, we use $\hat{R}=\sum_\ell (2\ell+1)(\hat{R}_\ell)_{ij}$ (note that this $\hat{R}$ is different from the residual $R^\beta_\nu$; see Sec.~\ref{sec.ILC}). Note that $g^{\beta'}_i$ is not $\ell$-dependent since we are fixing a fiducial $\beta'$ across all multipoles, and therefore, $d_i$ is also not $\ell$-dependent. The variance is thus
\begin{equation}
    \mathrm{Var}(y^{\beta'}_\alpha) = \frac{d_k (\hat{R}^{-1})_{ki}d_i}{(f_k (\hat{R}^{-1})_{ki}f_i) (d_m (\hat{R}^{-1})_{mn} d_n) - (f_k (\hat{R}^{-1})_{ki}d_i)^2} \, .
\end{equation}
Rather than minimizing the variance, we can also maximize the inverse variance:
\begin{equation}
    \label{eq.inv_var}
    (\mathrm{Var}(y^{\beta'}_\alpha))^{-1} = f_k (\hat{R}^{-1})_{ki}f_i-\frac{(f_k (\hat{R}^{-1})_{ki}d_i)^2}{d_k (\hat{R}^{-1})_{ki}d_i} \, .
\end{equation}
To do this in one pass is challenging because $\hat{R}$ depends on $h_\nu$, and $h_\nu$ depends on $\hat{R}$. We thus use an iterative process: 

\begin{enumerate}
    \item Start by fixing $h_\nu=0$, and compute $\hat{R}$.
    \item  Using that $\hat{R}$, find $h_\nu$ that maximizes Eq.~\eqref{eq.inv_var}. Specifically, using fixed $\hat{R}$, find the direction $v_\nu$ and scalar $\alpha$ that maximize the inverse variance in Eq.~\eqref{eq.inv_var} and compute $h_\nu=\alpha v_\nu/g^{\beta'}_\nu$. $\hat{R}$ is held fixed in this step to prevent the optimization from being noise-driven.
    \item  Use the new $h_\nu$ to compute $\hat{R}$, and repeat the process. We consider the process to have converged when the difference in $h_\nu$ from two rounds is less than some threshold, say 0.001.
\end{enumerate}

When $\hat{R}$ is fixed based on the value of $h_\nu$ from the previous round, the first term in Eq.~\eqref{eq.inv_var} is fixed. Thus, maximizing the inverse variance is equivalent to minimizing 
\begin{equation}
    \label{eq.quotient}
    \frac{(f_k (\hat{R}^{-1})_{ki}d_i)^2}{d_k (\hat{R}^{-1})_{ki}d_i} \, .
\end{equation}
This minimization makes sense with our objectives: 1) Minimizing the numerator corresponds to making $d_\nu$ close to orthogonal to $f_\nu$ and 2) Maximizing the denominator corresponds to $d_\nu$ being large in the directions of highest inverse covariance (largest signal-to-noise, with the CIB being the signal we want to inflate here). 

We define $v_\nu$ such that $h_\nu \equiv  \alpha v_\nu/g^{\beta'}_\nu$ so that $d_\nu = g^{\beta'}_\nu + \alpha v_\nu$. Then Eq.~\eqref{eq.orthogonality} becomes $\sum_\nu f_\nu^2 v_\nu / g^{\beta'}_\nu = 0 $. Minimizing the entire expression in Eq.~\eqref{eq.quotient} subject to the orthogonality constraint is challenging and computationally expensive, so instead we approximate the minimum using two steps: First, we find $v_\nu$ that maximizes the denominator while maintaining the orthogonality constraint (keeping the numerator fixed). Then, we find $\alpha$, related to the magnitude of $h_\nu$, that minimizes the full expression. This process gets repeated in each iteration.

To find $v_\nu$, we first project the inverse covariance $\hat{R}^{-1}$ into the subspace satisfying Eq.~\eqref{eq.orthogonality}. To project $\hat{R}^{-1}$ into this subspace, we compute the projection operator $P$ via $p_\nu = f_\nu^2/g^{\beta'}_\nu$ and $P=I-pp^T/(p^Tp)$. Then the projection is given by $P\hat{R}^{-1}P$. We then find the top eigenvector (the eigenvector corresponding to the maximum eigenvalue) of $P\hat{R}^{-1}P$ as $v_\nu$. This step maximizes the denominator in Eq.~\eqref{eq.quotient} (aside from an overall magnitude that is set by $\alpha$), subject to $v_\nu$ being orthogonal to $f_\nu^2/g_\nu^{\beta'}$. Next, we can analytically find $\alpha$ that minimizes Eq.~\eqref{eq.quotient}, given the $v_\nu$ we just found. Having found both $\alpha$ and $v_\nu$, we have solved for $h_\nu = \alpha v_\nu / g^{\beta'}_\nu $ for this iteration. We build new maps $T_\nu'$ with this $h_\nu$, solve again for the covariance $\hat{R}$, and repeat the procedure. When performing this process, we use only one fixed fiducial value $\beta'$ for determining $h_\nu$, and apply that $h_\nu$ to the residual for all values of $\beta$. This allows an apples-to-apples comparison of CIB leakage for all the tested $\beta$ values.

There are a few numerical tricks that speed up and stabilize convergence. To avoid rapidly oscillating values of $\alpha$ and $v_\nu$, we apply a technique similar to the idea of momentum in gradient descent. Letting $\alpha^{(i)}$ and $v_\nu^{(i)}$ denote the value of $\alpha$ and $v_\nu$ (respectively) at the $i$th iteration, we set
\begin{align}
    \alpha^{(i+1)} &= \alpha^{(i)} + \gamma (\alpha^{\rm candidate}-\alpha^{(i)}) \\
        v_\nu^{(i+1)} &= v_\nu^{(i)} + \gamma (v_\nu^{\rm candidate}-v_\nu^{(i)}) \, ,
\end{align}
where $\gamma$ is some constant less than 1 (set to 0.4 here), $\alpha^{\rm candidate}$ is the proposed new value of $\alpha$ based on the optimization of Eq.~\eqref{eq.quotient} at iteration $(i+1)$, and $v_\nu^{\rm candidate}$ is the proposed new value of $v_\nu$ at iteration $(i+1)$. We also add very small numbers to denominators of fractions to avoid numerical divergence.

\begin{figure}[t]
    \centering
    \includegraphics[width=1.0\textwidth]{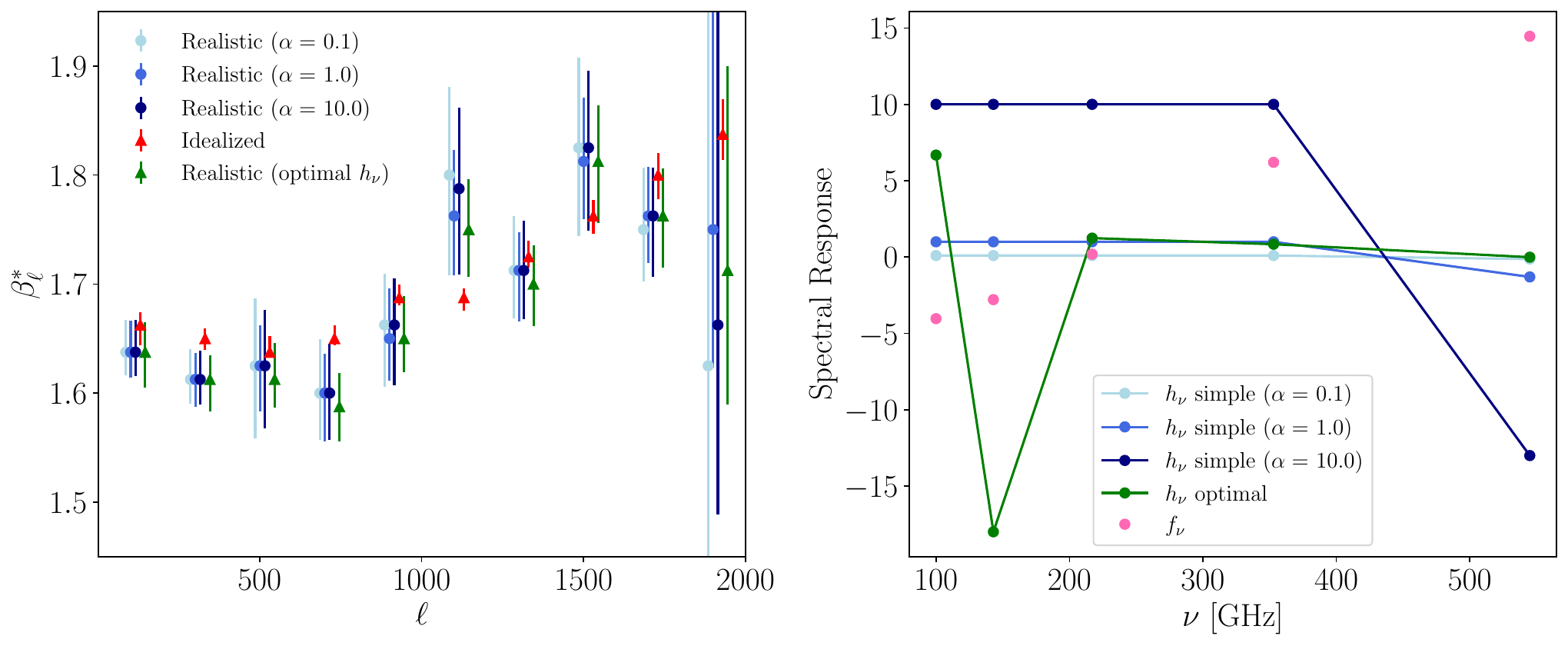}
    \caption{\textbf{Left:} Same as Fig.~\ref{fig:beta_vs_ell} but including $\beta^*_\ell$ for the case when $h_\nu$ is optimized (green). \textbf{Right:} Comparison of $h_\nu$ obtained using the simple parametrization from Sec.~\ref{sec.algorithm} for various values of $\alpha$ (shades of blue) to the optimal $h_\nu$ obtained via the approach detailed in this appendix (green). The \emph{Planck} bandpass-integrated tSZ spectral response $f_\nu$ is also shown in pink. Note that $h_\nu$ is dimensionless while $f_\nu$ has units of K.}
    \label{fig:beta_vs_ell_opt}
\end{figure}

With optimization of the $h_\nu$ selection in this manner, we find an average 15\% reduction in error bars on $\beta^*_\ell$ when using this method on the simulations described in Sec.~\ref{sec.simulations} with the baseline halo sample ($0.8 < z < 1.8$ and $10^{12}M_{\odot} < M < 10^{15}M_{\odot}$), as compared with setting $\alpha=1$ and using the $h_\nu$ parametrization described in Sec.~\ref{sec.algorithm}. With this new method we obtain $h_\nu = [6.679, -17.986, 1.242, 0.843, -0.007]$ (shown in Fig.~\ref{fig:beta_vs_ell_opt} and compared to $h_\nu$ obtained via the parametrization in Sec.~\ref{sec.algorithm}). The resulting $\beta^*_\ell$ are also shown in Fig.~\ref{fig:beta_vs_ell_opt}. While the error bars on $\beta^*_\ell$ shrink with this optimized way of determining $h_\nu$, the central values remain largely unchanged, with an average absolute shift of only half a percent (corresponding to a mean absolute shift of less than 0.01 in $\beta^*_\ell$). Because of our procedure of deprojecting only the central value $\beta^*_\ell$ in each bin, this means that there is practically no effect of this optimized $h_\nu$ selection on the final cross-correlation or its signal-to-noise ratio. Thus, this $h_\nu$ optimization would primarily be useful if one wanted to propagate the errors on $\beta^*_\ell$ into the error on the final cross-correlation measurement, though this is not what is done here. 

These findings are in line with expectations that using a suboptimal $h_\nu$ should not bias the final $\beta^*_\ell$ (so long as $\sum_\nu f_\nu^2h_\nu=0$ is satisfied), but do affect the error bars on $\beta^*_\ell$. The findings are also in line with what is seen in Fig.~\ref{fig:beta_vs_ell}, where different values of $\alpha$ all give consistent central values of $\beta^*_\ell$ but with slightly different error bars.


\section{Effects of Varying Halo Selection}
\label{app:diff_halos}

\begin{figure*}[t]
    \centering
    \includegraphics[width=0.9\textwidth]{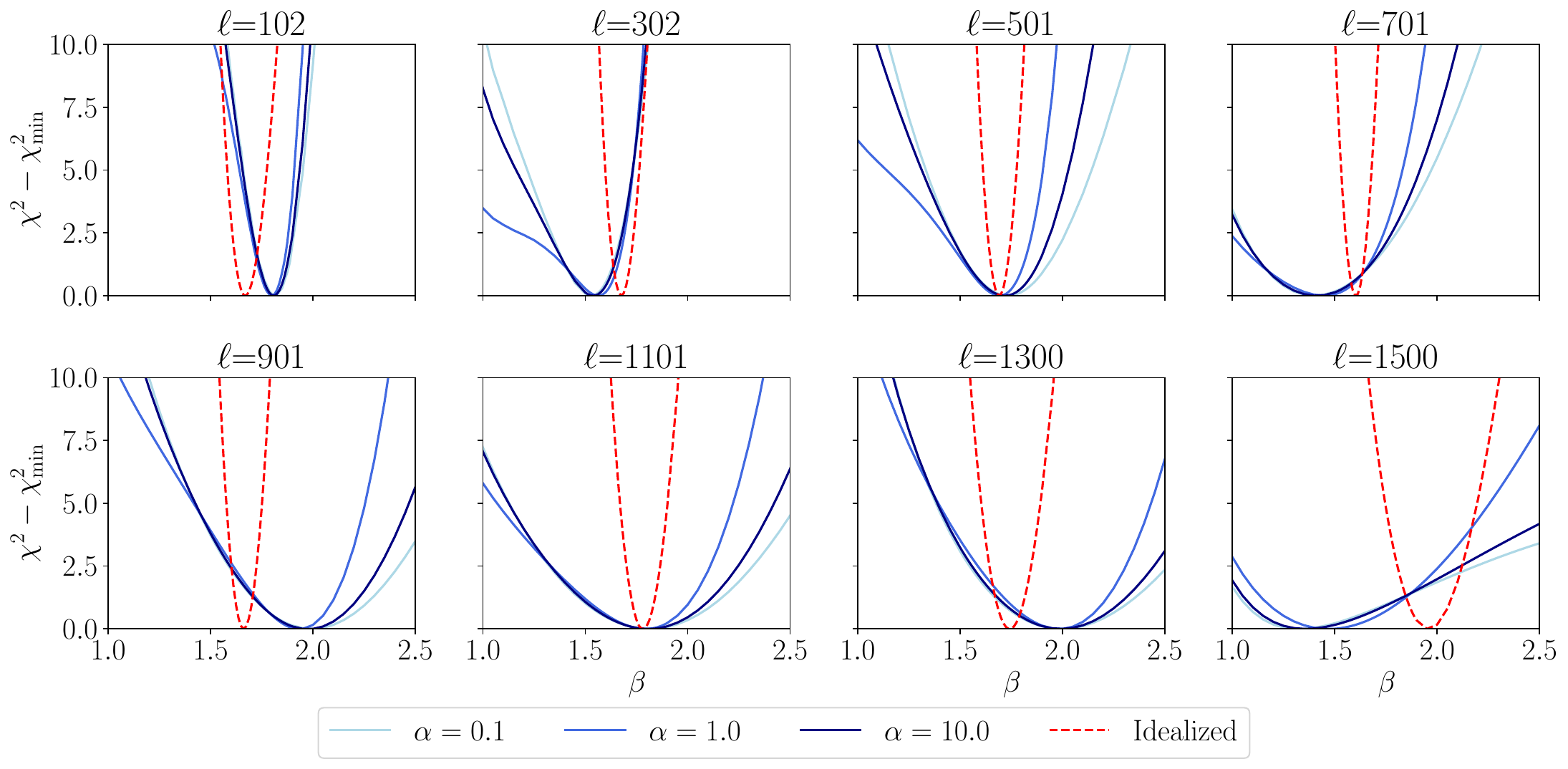}
    \caption{Same as Fig.~\ref{fig:chi2_vs_beta} but using halo selection of $0.3 < z < 0.8$ and $5 \times 10^{12} M_{\odot} < M < 5 \times 10^{14} M_{\odot}$.}
    \label{fig:chi2_vs_beta_midz}
\end{figure*}

\begin{figure}[t]
    \centering
    \includegraphics[width=0.48\textwidth]{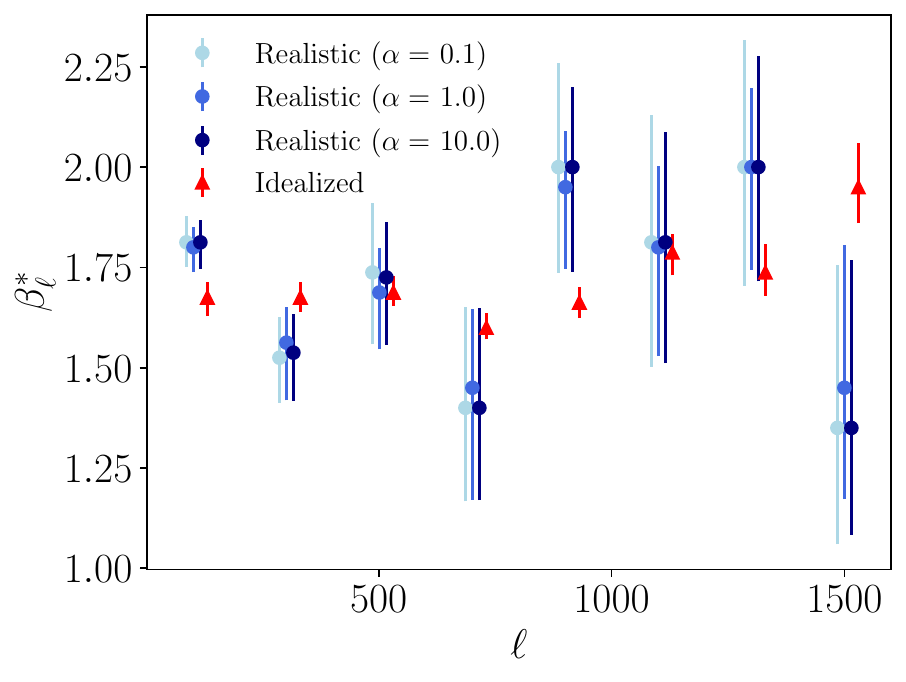}
    \caption{Same as Fig.~\ref{fig:beta_vs_ell} but using halo selection $0.3 < z < 0.8$ and $5 \times 10^{12} M_{\odot} < M < 5 \times 10^{14} M_{\odot}$.}
    \label{fig:beta_vs_ell_midz}
\end{figure}

In the main text, we showed how the algorithm works using one halo sample. Here, we show that the functionality of the procedure is not dependent on that specific halo selection, repeating the entire analysis on a different set of halos. In particular, here we consider halos in the redshift range $0.3 < z < 0.8$ and mass range $5 \times 10^{12} M_{\odot} < M < 5 \times 10^{14} M_{\odot}$. These halos are thus disjoint from those used in the main analysis, at lower redshifts. While the halo sample does not exactly correspond to a specific survey, it is roughly similar to, e.g., the DESI luminous red galaxies (LRGs), which have an approximate redshift range of $0.4 < z < 1.0$ \cite{DESI:2022gle}. Here we restrict our analysis to $2 \leq \ell \leq 1600$, a lower maximum $\ell$ than in the main analysis. This is because for this halo selection, in the \texttt{AGORA} simulations, the cross-correlation of the halo map and CIB at some frequencies becomes negative at lower $\ell$. We thus restrict the analysis to multipoles where the cross-correlation is positive.

Fig.~\ref{fig:chi2_vs_beta_midz} shows $\chi^2$ as a function of $\beta$ for each multipole bin using this halo selection, for various values of $\alpha$ (0.01, 1.0, and 10.0). Fig.~\ref{fig:beta_vs_ell_midz} shows the $\beta^{*}_\ell$ and $1\sigma$ error bars for each bin and value of $\alpha$. As with the main analysis, for every value of $\alpha$, $\beta^{*}_\ell$ agrees with the $\beta^*_\ell$ from the idealized pipeline within slightly over $1\sigma$.

\begin{figure*}[t]
    \centering
    \includegraphics[width=1.0\textwidth]{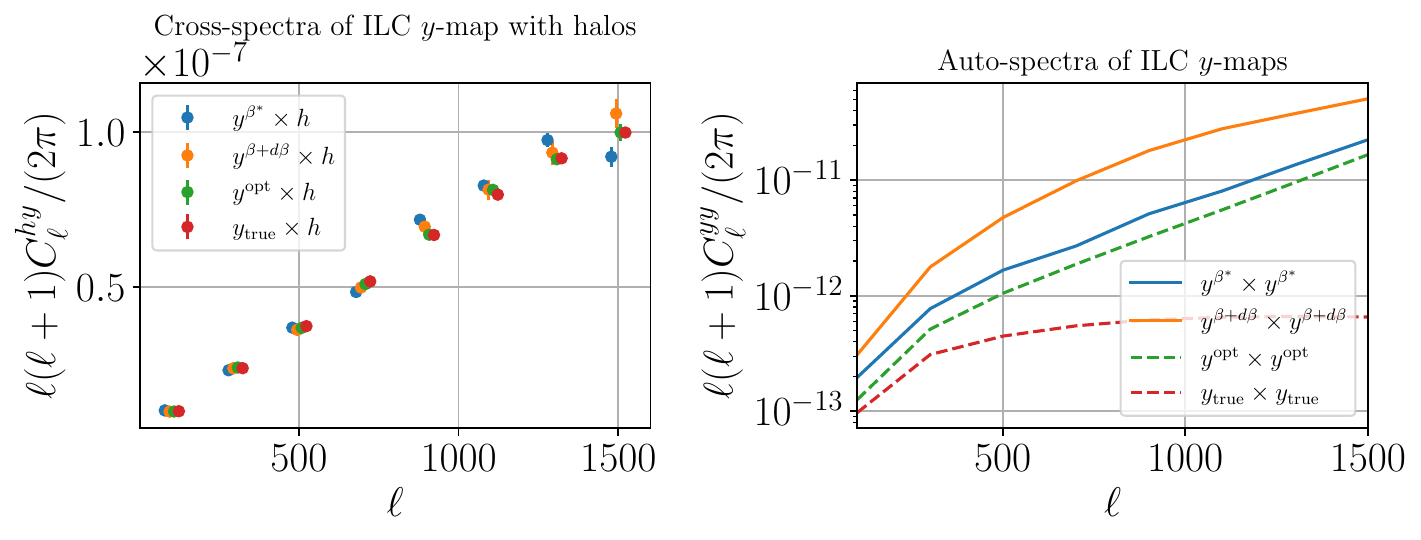}
    \caption{Same as Fig.~\ref{fig:dbeta} but using halo selection of $0.3 < z < 0.8$ and $5 \times 10^{12} M_{\odot} < M < 5 \times 10^{14} M_{\odot}$.}
    \label{fig:dbeta_midz}
\end{figure*}

\begin{figure}[t]
    \centering
    \includegraphics[width=0.4\textwidth]{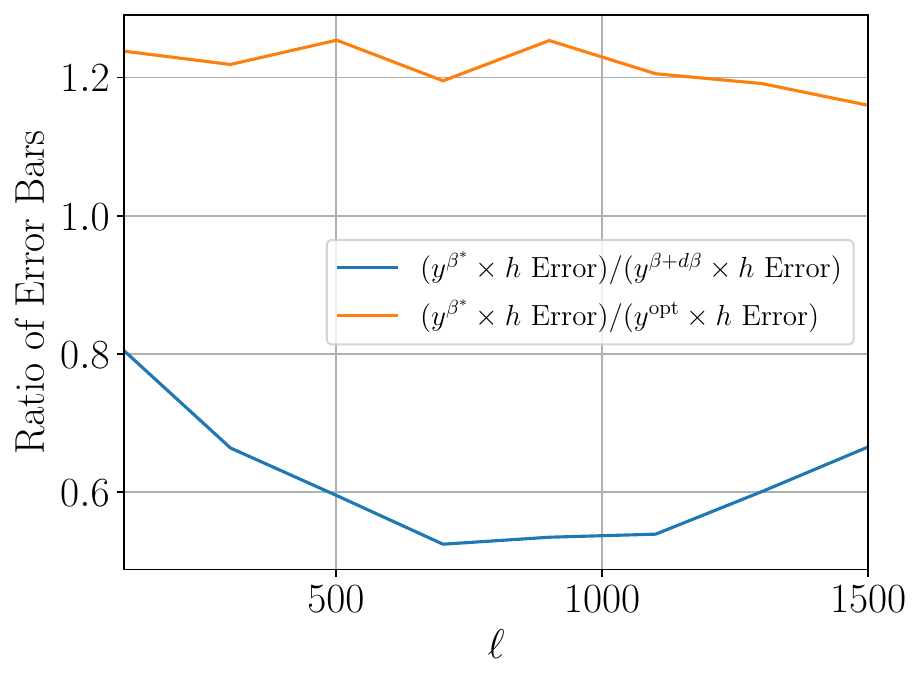}
    \caption{Same as Fig.~\ref{fig:error_ratio} but using halo selection of $0.3 < z < 0.8$ and $5 \times 10^{12} M_{\odot} < M < 5 \times 10^{14} M_{\odot}$. }
    \label{fig:error_ratio_midz}
\end{figure}

Fig.~\ref{fig:dbeta_midz} compares the resulting $y$-map, $y^{\beta^*}$, with one that includes moment deprojection, $y^{\beta + d\beta}$. Fig.~\ref{fig:error_ratio_midz} compares the size of the error bars from the $y$-maps. In terms of signal-to-noise, $y^{\beta^*} \times h$, $y^{\beta + d\beta} \times h$, and $y^{\rm opt} \times h$ have signal-to-noise ratios of 132, 80, and 160, respectively.

\section{CIB with No Decorrelation}
\label{app:no_decorr}

In this appendix, we consider artificially constructed simulations where the CIB does not decorrelate across frequencies. We construct an MBB SED using $T_d=24.0$ K and $\beta=1.65$. We take the 353 GHz CIB map as a fiducial CIB template, and apply appropriate factors of the SED to obtain the other frequency maps. In this case, we can check whether our recovered $\beta^{*}_\ell$ exactly matches the true $\beta=1.65$ that is used in the simulation construction. 

\begin{figure*}[htb]
    \centering
    \includegraphics[width=0.9\textwidth]{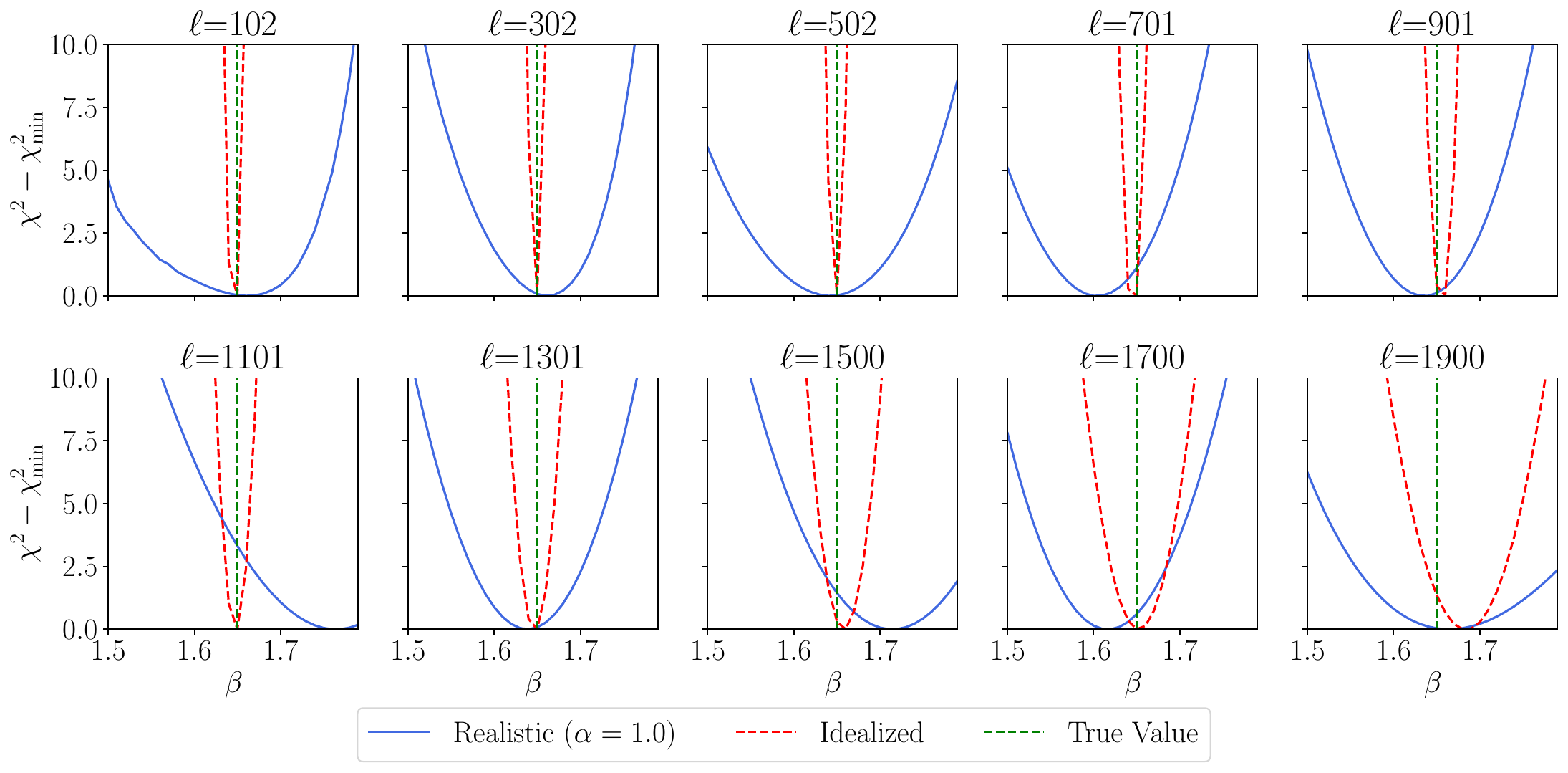}
    \caption{Same as Fig.~\ref{fig:chi2_vs_beta} but using simulations with no CIB decorrelation, i.e., where the CIB is perfectly described by an MBB with $\beta=1.65$. The dashed green line shows the true $\beta=1.65$ that is used in the simulation construction.}
    \label{fig:chi2_nodecorr}
\end{figure*}

\begin{figure}[htb]
    \centering
    \includegraphics[width=0.48\textwidth]{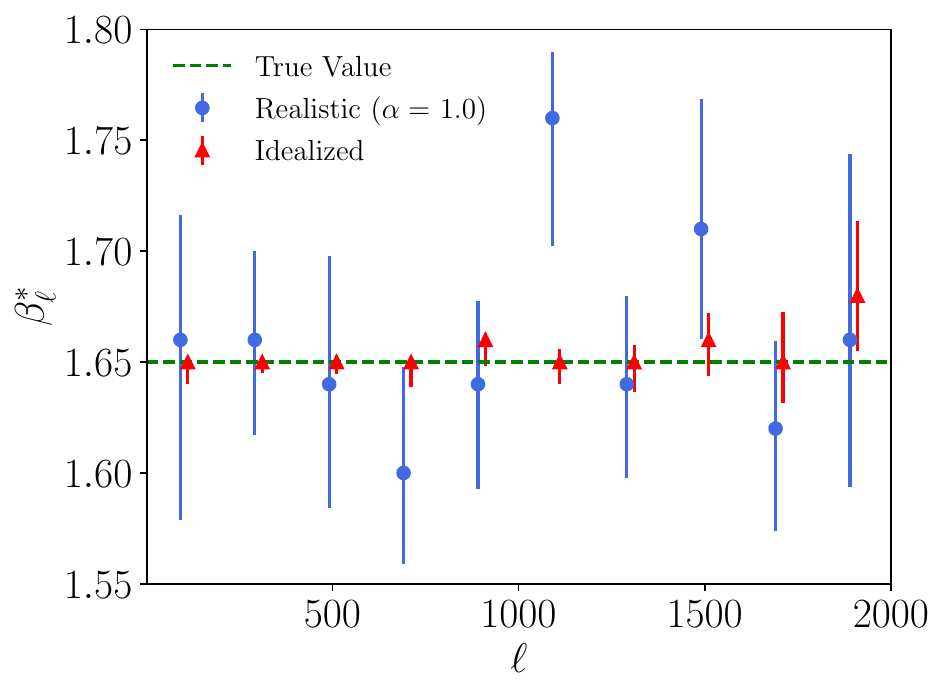}
    \caption{Same as Fig.~\ref{fig:beta_vs_ell} but using simulations with no CIB decorrelation, i.e., where the CIB is perfectly described by an MBB with $\beta=1.65$. The horizontal dashed green line shows the true $\beta$ value of 1.65 that is used in the simulation construction.}
    \label{fig:beta_vs_ell_nodecorr}
\end{figure}

For this test, we use $\alpha=1$ (though as previously shown, any value of $\alpha$ should work). We use the known $\beta=1.65$ for computation of the covariance matrix (although changing this has negligible impact on the final results). Moreover, we use the baseline halo selection of $0.8 < z < 1.8$ and $10^{12} M_{\odot} < M < 10^{15} M_{\odot}$. Fig.~\ref{fig:chi2_nodecorr} shows $\chi^2$ as a function of $\beta$, and Fig.~\ref{fig:beta_vs_ell_nodecorr} shows the optimal $\beta$ values obtained in each bin for these simulations. As expected, we recover $\beta^{*}_\ell=1.65$ sufficiently well given the error bars. 

\begin{figure*}[htb]
    \centering
    \includegraphics[width=0.85\textwidth]{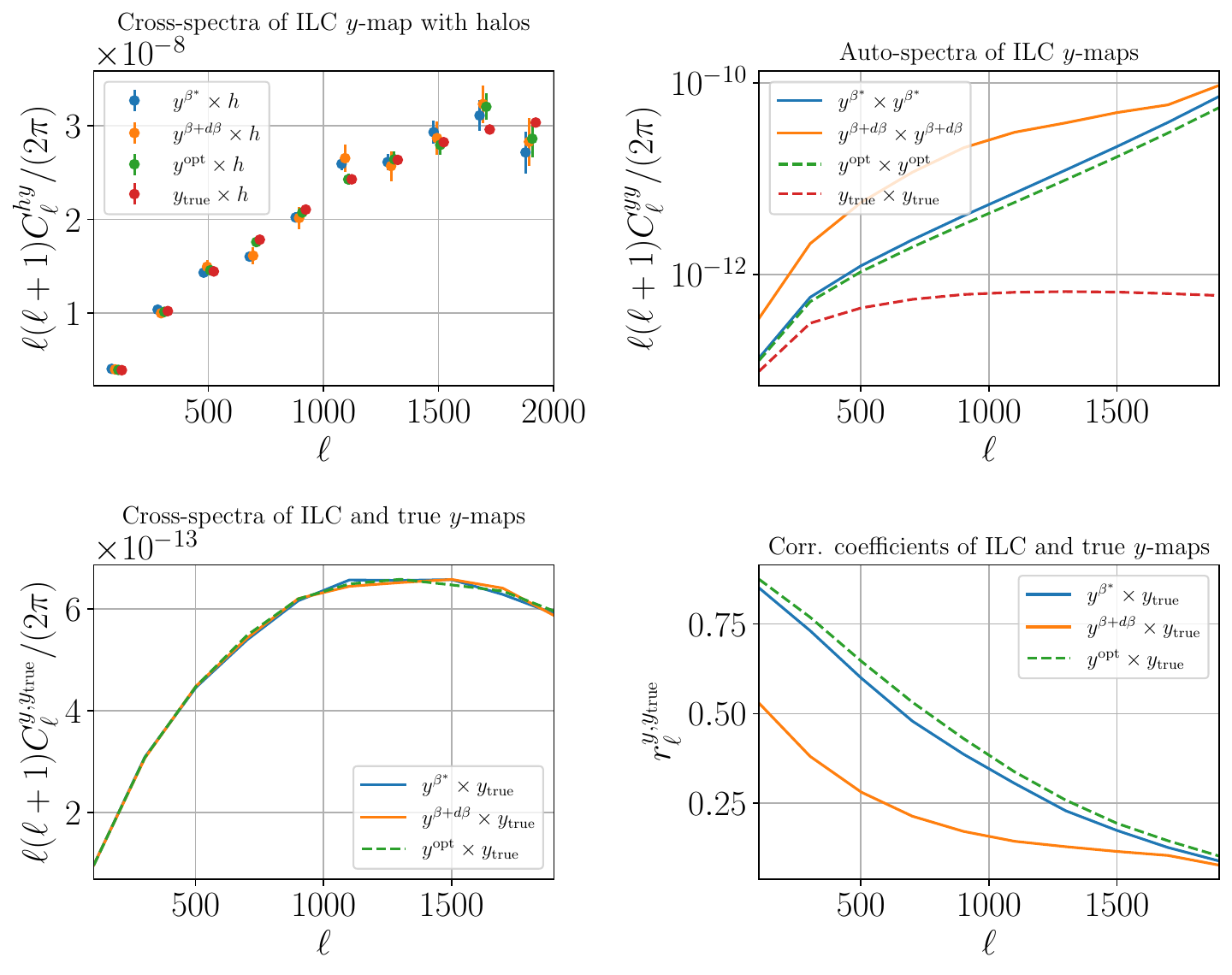}
    \caption{Comparison of our new method to ILC techniques that deproject both $\beta$ and $d\beta$. The map constructed from our new algorithm is denoted $y^{\beta^*}$ (blue). The map with moment deprojection is denoted $y^{\beta + d\beta}$ (orange). The standard ILC map built from frequency maps with noise but without CIB is denoted $y^{\rm opt}$ (green). The true $y$-map is denoted $y_{\rm true}$ (red). We note that only $y^{\beta^*}$ (blue) and $y^{\beta + d\beta}$ (orange) can actually be constructed from data. The other points are shown just for comparison. Here we use the baseline set of halos used in the main text, and simulations with no CIB decorrelation. \textbf{Top left:} Cross-spectra of ILC $y$-maps with halos, with Gaussian error bars. The differently colored points have slight horizontal offsets for visual clarity.  \textbf{Top right:} Auto-spectra of ILC $y$-maps. \textbf{Bottom left:} Cross-spectra of ILC $y$-maps with the true $y$-map. \textbf{Bottom right:} Correlation coefficients of ILC $y$-maps with the true $y$-map. }
    \label{fig:dbeta_nodecorr}
\end{figure*}

Fig.~\ref{fig:dbeta_nodecorr} shows a comparison of $y^{\beta^*}$ and $y^{\beta + d\beta}$ on these simulations that have no CIB decorrelation. Here, the central values of $y \times h$ from the two approaches are almost identical, but the auto-spectrum of $y^{\beta^*}$ remains significantly lower than that of $y^{\beta + d\beta}$. Interestingly, in this situation the cross-spectrum of $y^\beta \times y_{\rm true}$ matches that of $y^{\rm opt} \times y_{\rm true}$ and $y^{\beta + d\beta} \times y_{\rm true}$. This was not the case in situations with realistic CIB decorrelation across frequencies. Here this is possible because there is one true $\beta$ value that is used to produce the simulations, and thus it is theoretically possible to recover a single ``true" value of $\beta$. In the case of having decorrelation, there is no ``true" $\beta$ value, and so we optimize finding $\beta$ for a specific cross-correlation. 

\begin{figure}[htb]
    \centering
    \includegraphics[width=0.4\textwidth]{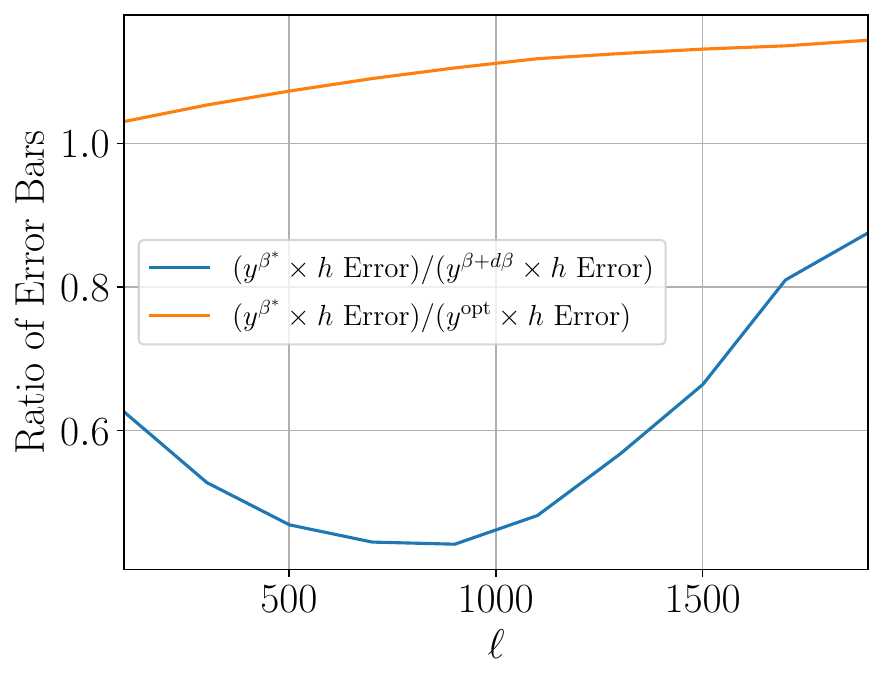}
    \caption{Same as Fig.~\ref{fig:error_ratio} but using simulations with no CIB decorrelation, i.e., where the CIB is perfectly described by an MBB with $\beta=1.65$.  }
    \label{fig:error_ratio_nodecorr}
\end{figure}

Fig.~\ref{fig:error_ratio_nodecorr} compares the size of the error bars from the $y$-maps. In terms of signal-to-noise, $y^{\beta^*} \times h$, $y^{\beta + d\beta} \times h$, and $y^{\rm opt} \times h$ have a signal-to-noise of 106, 57, and 117, respectively.

\bibliography{refs}

\end{document}